\documentclass[
reprint, 
twocolumn, 
secnumarabic,
tightenlines,
amssymb,
amsmath,
nobibnotes,
nofootinbib, 
floatfix,
aps, prd, 
showpacs,showkeys,
graphicx
]{revtex4-2}
\usepackage[draft]{hyperref}
\usepackage{cleveref}

\usepackage{multirow}%
\usepackage[caption=false]{subfig}
\expandafter \ifx \csname package@font\endcsname\relax\else
\expandafter \expandafter \expandafter
\usepackage
\expandafter \expandafter
\expandafter{\csname package@font\endcsname}%
\fi
\usepackage{textcomp}
\usepackage{gensymb}

\usepackage{float} 

\newcommand{\nuwro}{\text{\sc NuWro}}

\usepackage{forest}


\begin{document}

\title{Generative adversarial neural networks for simulating  neutrino interactions}

\author{Jos\'e L. Bonilla}
\email{joseluis.bonillaramirez@uwr.edu.pl}

\author{Krzysztof M. Graczyk}
\email{krzysztof.graczyk@uwr.edu.pl}

\author{\\Artur M. Ankowski}

\author{Rwik Dharmapal Banerjee}

\author{Beata E. Kowal}

\author{Hemant Prasad}

\author{Jan T. Sobczyk}

\affiliation{Institute for Theoretical Physics, University of Wroc\l aw, plac Maxa Borna 9,
50-204, Wroc\l aw, Poland}

\date{\today}%

\begin{abstract}
We propose a new approach to simulate neutrino scattering events as an alternative to the standard Monte Carlo generator approach. Generative adversarial neural network (GAN) models are developed to simulate charged current neutrino-carbon collisions in the few-GeV energy range.  We consider a simplified framework to generate muon kinematic variables, specifically its energy and scattering angle. GAN models are trained on simulation data from \nuwro{} Monte Carlo event generator. Two GAN models have been obtained: one simulating quasielastic neutrino-nucleus scatterings and another simulating all interactions at given neutrino energy. The models work for neutrino energy ranging from  300 MeV to 10 GeV.
The performance of both models has been assessed using two statistical metrics. It is shown that both GAN models successfully reproduce the distribution of muon kinematics.
\end{abstract}

\maketitle

\section{Introduction}

Monte Carlo (MC) event generators are indispensable tools in neutrino oscillation experimental analysis. Ongoing and future experimental collaborations, such as T2K~\cite{T2K:2019bcf}, MicroBooNE~\cite{MicroBooNE:2016pwy}, DUNE~\cite{DUNE:2020jqi}, and Hyper-Kamiokande~\cite{Hyper-KamiokandeProto-:2015xww}, use accelerator-based neutrino sources with broad neutrino energy spectra. Their goal is to perform precise measurements of the Standard Model (SM) oscillation parameters in the hope of gaining sensitivity to effects beyond the SM. However, the probability of neutrino oscillations is a rapidly changing function of neutrino energy, and its value cannot be directly accessed.

Thus, oscillation analyses require procedures to reconstruct neutrino energy from the detected final-state particles. These procedures rely heavily on MC simulations~\cite{doi:10.1146/annurev-nucl-102115-044720}, which must incorporate models for all relevant physical processes, including various types of interactions on bound nucleons (or, at higher energies, on quarks), contributions from two-body currents, and coherent neutrino-nucleus reactions. Additionally, final state interactions (FSI), i.e. secondary interactions of hadrons inside nuclei, must be accounted for.

Over the past $\sim 20$ years, significant efforts have been made to develop neutrino MC generators for modeling neutrino interactions with matter. Existing MC generators include those used directly by experimental groups, such as NEUT~\cite{Hayato:2021heg} and GENIE~\cite{Andreopoulos:2009rq}, as well as models developed by theoretical groups, such as GiBUU~\cite{Mosel:2018qmv}, NuWro~\cite{Golan:2012wx}, and Achilles~\cite{Isaacson:2022cwh}.

MC generators of events in particle and nuclear physics serve as a bridge between theory and experiment~\cite{campbell2024eventgeneratorshighenergyphysics}. Monte Carlo simulations are employed at every stage of an experimental study, including the production of the neutrino flux, interactions within a detector, and the detector's response. In this paper, we focus entirely on neutrino interactions inside the detector.

The MC generators, equipped with implementations of theoretical and phenomenological models, are often tuned to match experimental data. Theoretical models implemented in generators typically rely on uncertain parameters adjusted using available experimental data. However, these procedures cannot fully account for limitations stemming from theoretical models, which involve many approximations, particularly in the case of nuclear effects. Therefore, it would be beneficial to develop alternatives to existing MC event generators that can learn directly from available experimental cross section data and modify their predictions accordingly. As a result, efforts have been made to explore new approaches for developing alternatives to standard MC generators. A novel approach has recently emerged: leveraging advanced artificial intelligence (AI) models to dynamically adapt to experimental data, enabling highly effective simulations of physical phenomena.
Machine-learning methods have been applied to study and model neutrino-nucleus scattering and related electron-nucleus processes. These applications include extracting information about the axial content of the nucleon~\cite{Alvarez-Ruso:2018rdx}, modeling electron-nucleus scattering cross sections~\cite{AlHammal:2023svo,Kowal:2023dcq,graczyk2024electronnucleuscrosssectionstransfer,Sobczyk:2024ajv}, and accelerating sampling from given differential cross section distributions~\cite{ElBaz:2023ijr,baz2025efficientmontecarloevent}.

This paper explores, adapts, and advances modern machine-learning techniques for simulating neutrino-nucleus scattering events. We focus on generative adversarial networks (GANs)~\cite{goodfellow2014generativeadversarialnetworks}. The GAN technique enables training a neural network that learns to generate data produced by an unknown or poorly understood mechanism. A successful generative model takes as input a latent vector drawn from a given distribution in the latent space (typically a Gaussian distribution) and generates samples that match the characteristics of the training data.

Generative deep learning models have demonstrated their effectiveness across various domains, including image genera	tion~\cite{nguyen2017plugplaygenerative}, image super-resolution, and semi-supervised learning~~\cite{ledig2017photorealisticsingleimagesuperresolution,NIPS2016_8a3363ab}.  A comprehensive review of GANs can be found in Ref.~\cite{goodfellow2017nips2016tutorialgenerative}. The GAN technique is also promising for MC generators of particle interactions~\cite{Butter:2022rso}. Indeed, the GANs have been adapted to simulate parton showers~\cite{deOliveira:2017pjk,Monk:2018zsb} in high energy collisions, as well as to model hadronization~\cite{Ghosh:2022zdz,Ilten:2022jfm,Chan:2023ume}. An effort has also been made to adapt GANs to simulate collisions of highly energetic electrons with proton~\cite{Alanazi:2020jod}.

This paper applies the GAN technique to simulate charged current (CC) neutrino-carbon scattering. We adopt a simplified framework wherein the models are tasked with generating muon kinematic variables. We consider two cases. First, we test the applicability of our approach to quasielastic (QE) scattering. Then, we extend our study to include all interaction modes at a given neutrino energy. In both cases, the GAN takes neutrino energy as input and generates the charged lepton kinematics. The model is applicable to neutrino energies ranging from a few hundred MeV to tens of GeV, which are typical for neutrino oscillation experiments.

For our study, we utilize the \nuwro{} MC generator, which has been under development since 2004~\cite{Sobczyk:2004va,Juszczak:2005zs,Golan:2012rfa}. It is optimized for the energy range characteristic of neutrino oscillation experiments using accelerator-based neutrino sources, spanning from hundreds of MeV to tens of GeV. The GAN is trained on NuWro-generated data to produce scattering events.
Developing a successful GAN model is more challenging than standard supervised learning~\cite{Das:2023ktd}. Therefore, we assess the model’s performance using two metrics to evaluate its effectiveness: the mean absolute value of pulls and the Earth Mover’s Distance (EMD)~\cite{710701}.

The paper is organized as follows: Sec.~\ref{Sec:Method} introduces the GAN techniques and describes the technical details of our approach; Sec.~\ref{Sec:Numerical} presents numerical results; Sec.~\ref{Sec:Summary} summarizes the paper.

\section{Method}

\label{Sec:Method}

We aim to obtain a deep neural network (DNN) that generates neutrino-nucleus scattering events. As an illustrative example, we consider muon neutrino charged current interactions with a carbon target. The DNN model is trained (optimized) to learn the underlying probability distribution. As a~source of information on neutrino-carbon scattering, we employ the \nuwro{} MC generator.

In realistic simulations, one provides the input information and models the neutrino-nucleus collisions in all relevant channels. As an output, one obtains a distribution of events, including complete information about the final particles. The number of final particles can vary from event to event. Obtaining a generative network that does the same work as ``classical'' MC generators is a difficult task. Therefore,  in this first effort, we consider  a~GAN that generates only the final lepton's kinematics, and the flux spectrum is uniform.  

The procedure for developing the model is the following: 
\begin{itemize}
    \item Define the lepton variables generated by the model;
    \item Obtain the training dataset; 
    \item Construct GANs for analysis;
    \item Perform optimization; 
    \item Test model against data not used in the optimization stage. 
\end{itemize}
Each step is described in the following subsections.

\subsection{Data and kinematic settings}

\begin{figure}[htbp]
    \includegraphics[width=0.45\textwidth]{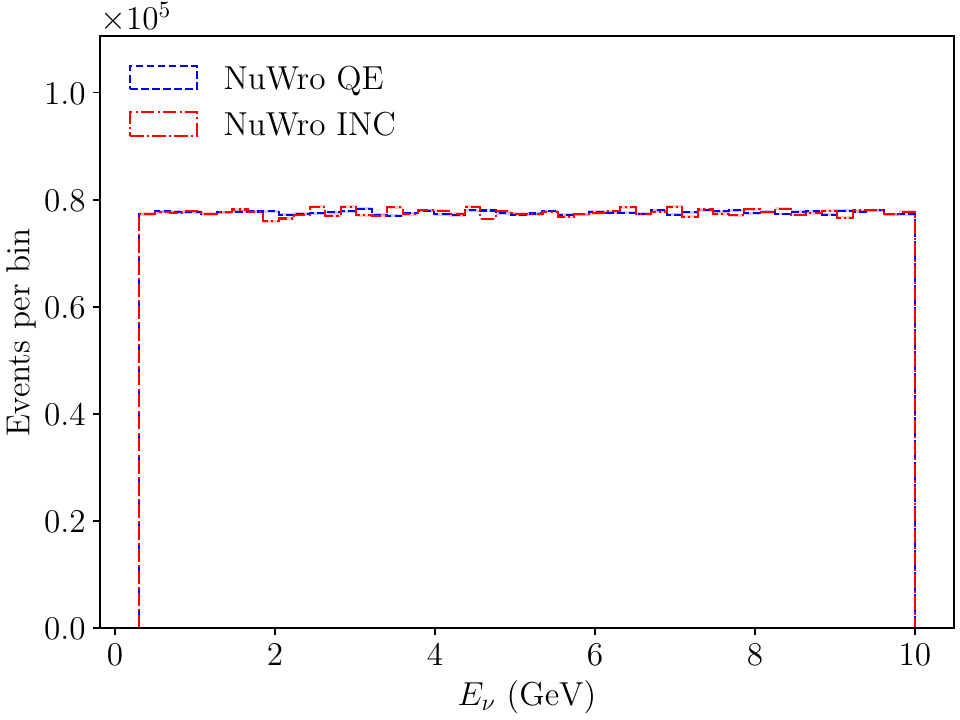}
    \caption{Energy distribution of the QE (blue dashed line) and INC  (red dashed-dotted line) training samples from \nuwro{}.
    \label{Fig:Energy-spectrum_QE}}
\end{figure}

We consider CC QE  and inclusive (INC) scattering of muon neutrinos off carbon.  The QE and INC data are obtained from a simulation done with the \nuwro{} MC event generator, using the spectral function approach to model the nucleus in the QE channel~\cite{Benhar:1994hw} and the local Fermi gas in other channels.
In the QE case, the generated events contain muon and nucleons or sometimes pions due to the final state interaction effects. In the INC case, the final states contain muons and a variety of mesons and baryons. In both types of events, this study only considers the muon's kinematics.

Let us denote the neutrino energy by $E_\nu$. The outgoing muon is characterized by energy $E_\mu$ and momentum $\mathbf{p}_\mu$. The angle between the neutrino and muon momenta is denoted by $\theta$. 

The complete description of muon kinematics requires three independent variables, such as the components of the muon momentum. However, due to the rotational symmetry of the system relative to the axis defined by the neutrino beam, two independent variables are sufficient to generate the muon kinematics for unpolarized targets. These could be chosen: the muon energy $E_\mu$ and the longitudinal momentum $p_{\mu,z}$ (along the neutrino momentum) or $E_\mu$ and the angle $\theta$. 

The variables $p_{\mu,z}$, $E_\mu$, and $\theta$ are related as follows:
\begin{equation}
    p_{\mu,z} = \sqrt{E_\mu^2 - m_\mu^2}\cos\theta,
    \label{Eq:pmuz}
\end{equation}
where $m_\mu$ is the muon mass.

Our goal is to construct a model that, for a given energy $E_\nu$, generates a required distribution of the events. We consider neutrino energies  from $E_{\nu, min} = 300$~MeV to $E_{\nu, max} = 10$~GeV. 

The neural networks work most efficiently when input variables are in the range $\sim(-1,1)$. Hence, we introduce the rescaled energy variable:
\begin{equation}
\label{Eq:Eprime}
	E_{\nu}^\prime = 2 \frac{E_\nu - E_{\nu, min}}{E_{\nu, max} - E_{\nu, min}} - 1,
\end{equation}
then $E'_\nu \in [-1,1]$.

We optimize the GAN model on events with uniform energy distribution to obtain a model that accurately predicts neutrino events for any energy in the considered range. We generate a large training set such that $\sim$40k events correspond to any energy interval of $100$~MeV, with a total of $3.88$M events comprising each training sample. The energy distributions of the events used to train both the QE and INC models are shown in Fig.~\ref{Fig:Energy-spectrum_QE}.

As explained above, the network generates two variables that define the muon kinematics. After considering various options, we  use a pair $(E_\mu^\prime, \theta')$, given by\footnote{We explored various transformations of the kinematic variables
to optimize the performance of our neural network model. We aimed to transform
the data into a format where the model operates most efficiently and where
optimization yields the best results. We chose a range of [-1, 1] because in
this domain, typical activation functions perform most effectively. Additionally,
we applied a square root transformation to reduce skewness in the event
distributions, which aids the training process.}:
\begin{eqnarray}
\label{Eq:Emuprime}
  E_\mu^\prime & = & 2\sqrt{1-\frac{E_\mu-m_\mu}{\Delta E}} - 1, \quad \Delta E = E^-_\nu - m_\mu \\
 \theta^\prime & = & 2\sqrt{\frac{\theta}{\pi}} - 1, 
 \label{Eq:thetaprime}
\end{eqnarray}
where $ E^-_\nu =  E_\nu - 9.8$~MeV is used to ensure a tighter upper bound\footnote{It effectively takes into account the effect of binding energy.}.
Both variables take values in the range approximately from $-1$ to $1$.

\subsection{GAN for event generation}

The objective of the GAN approach is to develop a model (usually DNN) such that, given a random latent vector, it predicts outputs that belong to the ``true''\footnote{In the case of present studies ``true'' denotes  \nuwro{} generated event.} distribution of samples. In this context, we are using a conditional GAN (CGAN)~\cite{DBLP:journals/corr/MirzaO14}, which incorporates neutrino energy and the random latent vector.

\begin{figure}[hbtp]
    \subfloat[\label{fig:block1}]{\includegraphics[scale=1.5]{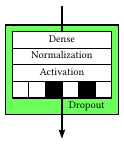}} \ \
    \subfloat[\label{fig:block2}]{\includegraphics[scale=1.5]{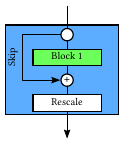}}
    \caption{\protect Building blocks of the generators and discriminators, \texttt{Block 1} (a) and \texttt{Block 2} (b).  The second block is obtained from \texttt{Block 1} by adding a skip connection and rescaling layer.
    \label{fig:blockschematics}}
\end{figure}

\begin{figure}[hbtp]
    \centering
    \includegraphics[scale=1.5]{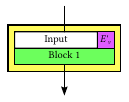}
    \caption{The condition $E'_\nu$ is concatenated to the inputs of a block of the INC network. The \texttt{Block 1} (see Fig. \protect\ref{fig:blockschematics}) with above modification is called \texttt{Block 3}. If we add a skip connection and rescale layer to a \texttt{Block 3},  one obtains \texttt{Block 4}.
    \label{fig:block3}}
\end{figure}

To obtain a successful generator G, one introduces a discriminator D, another DNN trained to discriminate between samples obtained from the generator and the ``true'' samples from the training dataset. Both generator and discriminator models are optimized. The discriminator learns to distinguish between ``true'' and GAN events, while the generator improves in  ``mimicking true'' events discriminator.

We implement models and perform numerical analysis using the Keras 2.15 package~\cite{chollet2015keras}. Note that the architectures of both QE and INC models are similar. Let us denote the discriminator and generative models for QE(INC) data by D-QE(INC) and G-QE(INC), respectively.

The common elements of the basic building block are shown in Fig.~\ref{fig:blockschematics}. The basic module, \texttt{Block 1}, of our DNNs consists of a fully connected (dense) layer, followed by a normalization~\cite{ba2016layernormalization}, an activation layer, and a dropout layer (with a rate of $10\%$); see Fig.~\ref{fig:blockschematics}(a). 
If the input and output of the \texttt{Block 1} in the main body of the network have the same dimension, a skip connection~\cite{NEURIPS2018_a41b3bb3} from the input to the output is added (see Fig.~\ref{fig:blockschematics}(b)) to form \texttt{Block 2}. The output of this block is rescaled by factors we discuss below.

We also employ a variation of \texttt{Block 1 (2)}, called \texttt{Block 3 (4)}, in which the information about the condition $E'_\nu$ is concatenated to its input right before the dense layer, as shown in Fig.~\ref{fig:block3}. \texttt{Block 1(2)} with extra $E'_\nu$ input will  be called \texttt{Block 3(4)}, respectively. Note that these blocks are used only by INC models\footnote{The inclusive data is more complex than the QE data.
Initially, we considered using the same neural network model to generate both
INC and QE events. However, we could not get satisfactory results for the INC data, and we ultimately decided to implement some modifications to better capture the
more intricate structure of the data.}.

\begin{figure}[htbp]\centering
    \subfloat[\label{fig:geninput}]{\includegraphics[scale=1.5]{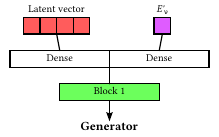}} \\
    \subfloat[\label{fig:disinput}]{\includegraphics[scale=1.5]{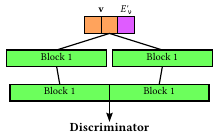}}
    \caption{The input block of the generators (a) and discriminators (b). The generator independently transforms the latent vector and $E'_\nu$ by a linear dense layer. Then, both outputs are concatenated and sent to a \texttt{Block 1}. The discriminator directly concatenates $\mathbf{v}$ and $E'_\nu$, then processes them in two parallel branches with two consecutive blocks. \label{fig:NetworkInput}}
\end{figure}

The input latent vector $\mathbf{x}$ is drawn from the standard normal distribution $N(0,1)$. Generators G-QE and G-INC transform the input latent vector and neutrino energy similarly. Namely, both $\mathbf{x}$ and $E'_\nu$ are transformed independently by a fully connected dense layer (with a \texttt{linear} activation function) to ensure that the neutrino energy is as important as the latent vector; see Fig.~\ref{fig:NetworkInput}(a). Then, both layers are concatenated together, processed by \texttt{Block 1}, and sent to the \textit{main body} of the network. The main body of G-QE(INC) consists of one \texttt{Block 1(3)} followed by one \texttt{Block 2(4)}. Consequently, there are a total of three blocks chained in G-QE(INC). The last section consists of the dense layer with a two-dimensional output activated with \texttt{tanh} function. 
\begin{figure}[htbp]
    \begin{center}
    \includegraphics[width=0.48\textwidth]{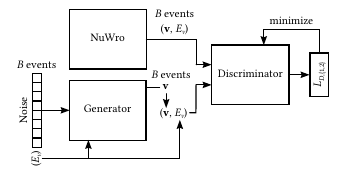}\\
    \includegraphics[width=0.48\textwidth]{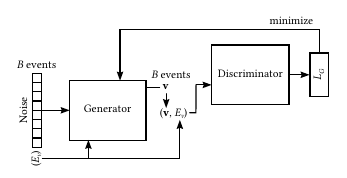}
    \end{center}
    \caption{The top graph shows the iteration step in which the discriminator trained to recognize the \nuwro{} and the network-generated samples. Discriminator takes for input $\mathbf{v} = (E'_\mu,\theta')$, and the condition $E'_\nu$. The bottom graph shows the iteration step in which the generator is optimized. \label{fig:gantraining}}
\end{figure}

The input of D-QE(INC) is either the \nuwro{} sample vector $\left(\mathbf{v}, E'_\nu\right)$ or the output of the generator $\left(G(\mathbf{x}, E'_\nu), E'_\nu\right)$. The input is transformed in parallel by two branches of the same architecture (see Fig.~\ref{fig:NetworkInput}(b)). Each branch consists of a sequence of two \texttt{Block 1} layers. Then, the outputs of both branches are concatenated and processed by the \textit{main body}, composed of one \texttt{Block 1(3)} and a sequence of three \texttt{Block 2(4)} layers. The neural network has a total depth of six \texttt{Block} layers. The final section of the discriminator is a fully connected layer with a single neuron and a \texttt{linear} activation function (a \texttt{sigmoid} is applied to the output when it contributes to the loss).

We apply \texttt{leaky ReLU}~\cite{Maas2013RectifierNI} activation functions for the D-QE and G-QE models. Since the INC samples are more complex than the QE ones, we improved the network architectures to converge with the optimization process. We observed that the discriminator has to be more robust than the generator to feed relevant information to the generator. With this in mind, the activation functions in the blocks were changed to \texttt{ReLU}~\cite{pmlr-v15-glorot11a} for G-INC and \texttt{PReLU}~\cite{2015arXiv150201852H} for D-INC. As remarked, \texttt{Blocks 1 and 2} of the central bodies of the INC networks are replaced by \texttt{Blocks 3 and 4}. The rescaling of \texttt{Blocks 4} is $1/2$ in the INC model, and there is no rescaling in \texttt{Blocks 2} for the QE model.

Additionally, to improve the training convergence, we supplement the structure of the network architectures by including Gaussian noise layers~\cite{pmlr-v139-feng21g}, defined by standard deviation $10^{-4}$ for G-QE and $10^{-3}$ for D-QE and D-INC. The G-INC is not affected by Gaussian noise. The inputs of G-QE, D-QE, and D-INC are injected with noise, as well as each block of G-QE and D-QE after each normalization layer, while the D-INC includes it after each layer with learnable parameters (dense, normalization layer, and \texttt{PReLU}). The last dense layer of G-QE is not affected, but the output of the last dense layer of both D-QE and D-INC are also altered by noise. It is worth noting that \texttt{Blocks 3} and \texttt{4} of D-INC are fed with the noisy $E'_\nu$.

The blocks of the G-QE and the D-QE have $100$ neurons, except for the first \texttt{Block 1}  of G-QE, which has $200$ neurons, and two dense layers of each initial pair of branches of the D-QE, which consists of $50$ neurons. Similarly, the G-INC comprises blocks and dense layers of $100$ neurons, but the first block has $200$ neurons (as the G-QE). The D-INC model has more neurons in the blocks than its QE counterpart. In D-INC, most blocks consist of $141$ neurons, but the first two dense layers in the initial branches have $70$ neurons each.

\subsection{Optimization scheme}

To train the model, we use the cross-entropy loss function given by three contributions
\begin{eqnarray}
\label{Eq:Dloss2}
L_{D,1}(\textbf{v}) &=& -\frac{1}{B}\sum_{j=1}^B \log[1-D(\textbf{v}_j, E_{\nu,j})], \ \\
\label{Eq:Dloss1}
L_{D,2}(G) &=& -\frac{1}{B}\sum_{i=1}^B \log [D(G(\mathbf{x}_i, E_{\nu,i}),E_{\nu,i})],
\\
\label{Eq:Gloss}
L_{G}(G) & = & -\frac{1}{B}\sum_{k=1}^B \log[1 - D(G(\mathbf{x}_k,E_{\nu,k}), E_{\nu,k})],
\end{eqnarray}
where $B$ is the number of samples in the mini-batch.

GANs are optimized using the so-called mini-max framework~\cite{Goodfellow-et-al-2016}. Hence, a successful model is a result of a balance between two losses: of the discriminator (\ref{Eq:Dloss2}) and (\ref{Eq:Dloss1}) and of the generator (\ref{Eq:Gloss}).
$L_{D,1}$ in Eq. (\ref{Eq:Dloss2}) is calculated on ``true'' samples obtained from \nuwro{}, whereas the loss (\ref{Eq:Dloss1}) is computed on ``fake'' data generated by G. The last loss (\ref{Eq:Gloss}) is evaluated on generated by model ``fake'' samples.

Both the generator and discriminator are optimized simultaneously in the minibatch configuration. In the single iteration step, the discriminator is optimized with minibatch for the \nuwro{} data consisting of \( B = 10,000 \) and \( B = 1,000 \) samples for the QE and INC analyses, respectively. In the same iteration step, the generator is optimized with a minibatch size of \( 2B \). The iterations are repeated until no data is left in the \nuwro{} training dataset, and a single epoch is accomplished. The training loop is shown in Fig.~\ref{fig:gantraining}. We utilize the \texttt{AdamW} optimization algorithm for this purpose, a learning rate of $10^{-4}$ for D-QE(INC) and $10^{-5}$ for G-QE(INC), and $\beta_1 = 0.5$ and $\beta_2 = 0.9$.

\subsection{Testing model's quality}

\begin{figure*}[htbp]\centering
    \includegraphics[width=0.45\textwidth]{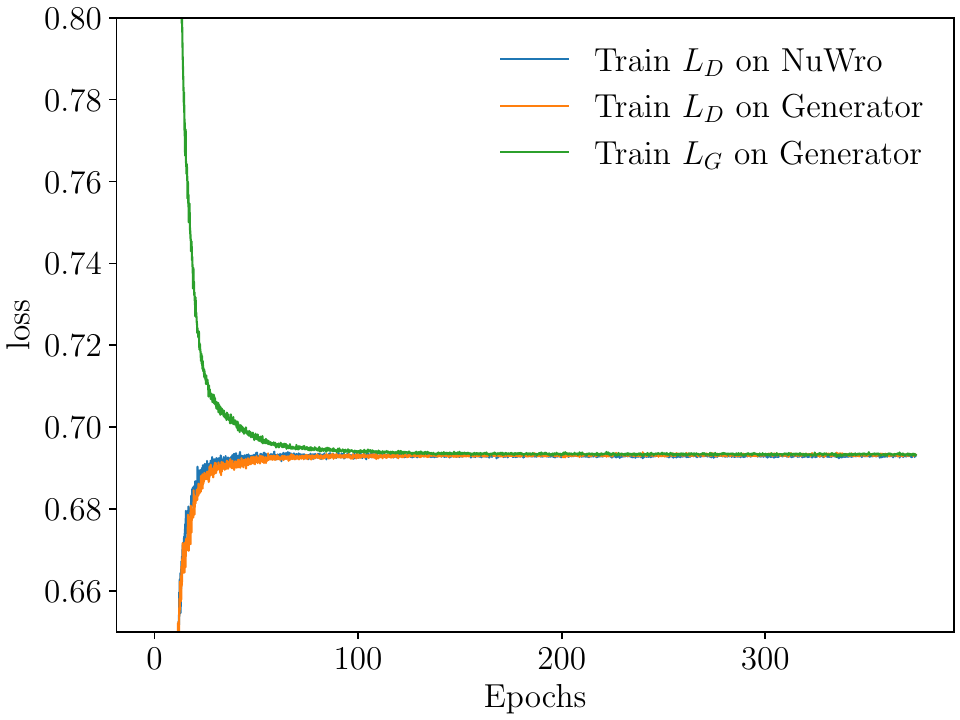} \ \ \
    \includegraphics[width=0.45\textwidth]{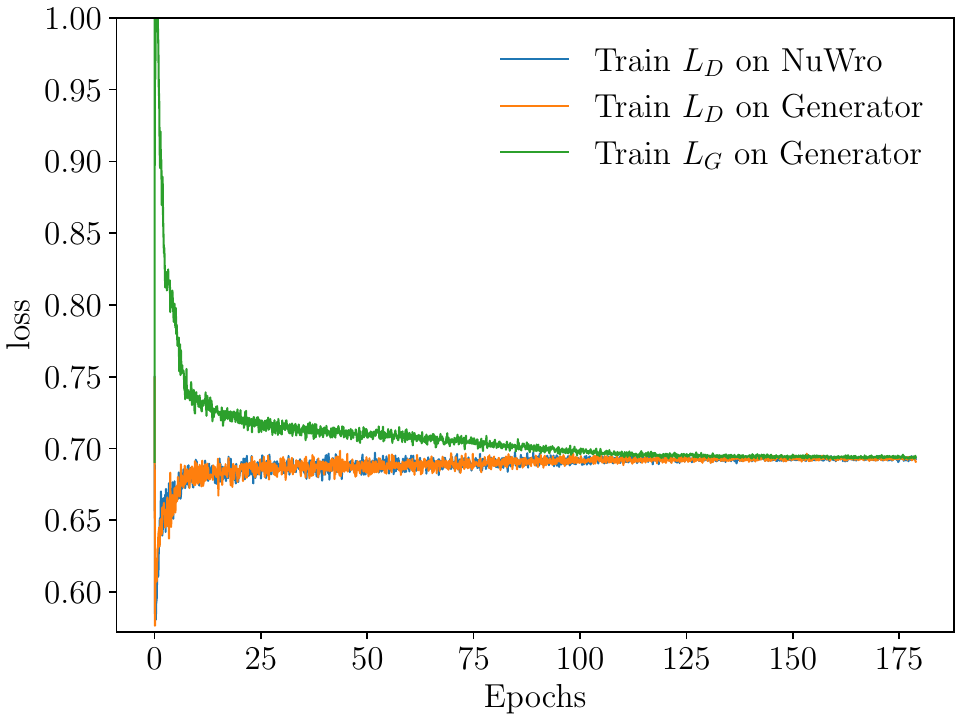}
    \caption{The losses (\protect\ref{Eq:Dloss1}), (\protect\ref{Eq:Dloss2}), and (\protect\ref{Eq:Gloss})  evaluated to monitor the training of the QE (left figure) and INC (right figure) GAN models.     \label{fig:performance-QE}}
\end{figure*}

We monitor the optimization of the generator and discriminator by evaluating their respective loss values.
In Fig.~\ref{fig:performance-QE}, we present the variation of the losses during the training of the GANs on the QE and INC data. In both cases, the loss curves converge to the same value, corresponding to the model configuration where a balance is achieved between the generator's and the discriminator's performance.

\begin{figure*}[htbp]
    \begin{center}
    \includegraphics[width=0.75\textwidth]{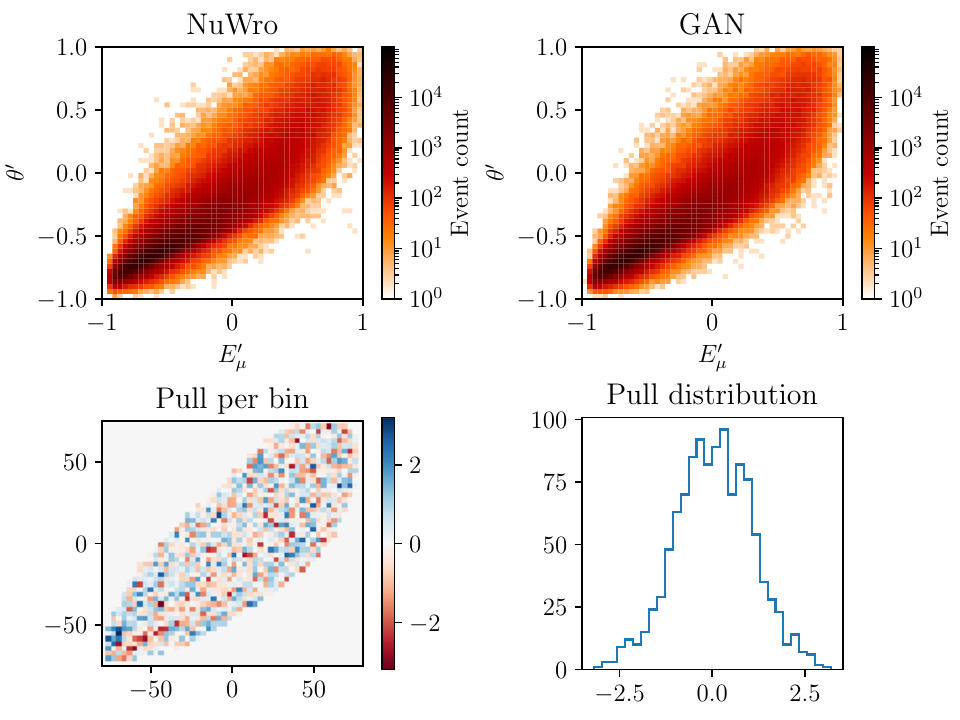} 
    \end{center}
    \caption{Top row: \nuwro{} test sample (left) and GAN (right) generated samples for the QE scattering. Bottom row: Pull w/o tails per bin from \nuwro{} and G-QE samples (left) and pull distribution (right). 
    \label{fig:pulls_map1}}
\end{figure*}

\begin{figure*}[htbp]
    \includegraphics[width=0.75\textwidth]{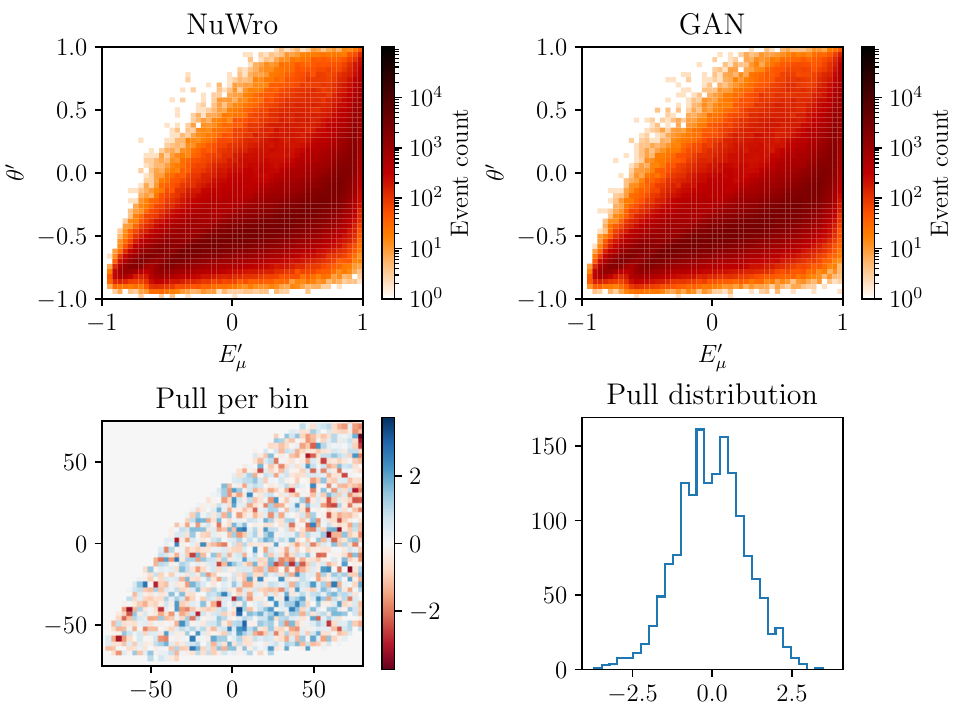} 
    \caption{Caption the same as in Fig.~\protect\ref{fig:pulls_map1} but for inclusive scattering.
    \label{fig:pulls_map2}}
\end{figure*}

\begin{figure*}[htbp]\centering
    \includegraphics[width=0.45\textwidth]{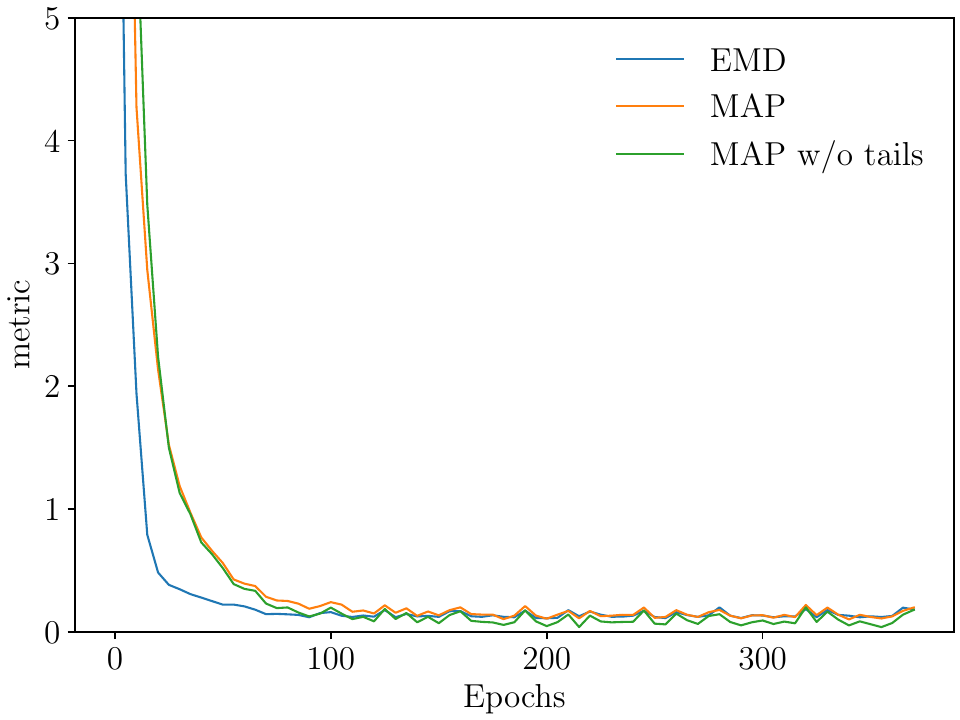} \ \ \
    \includegraphics[width=0.45\textwidth]{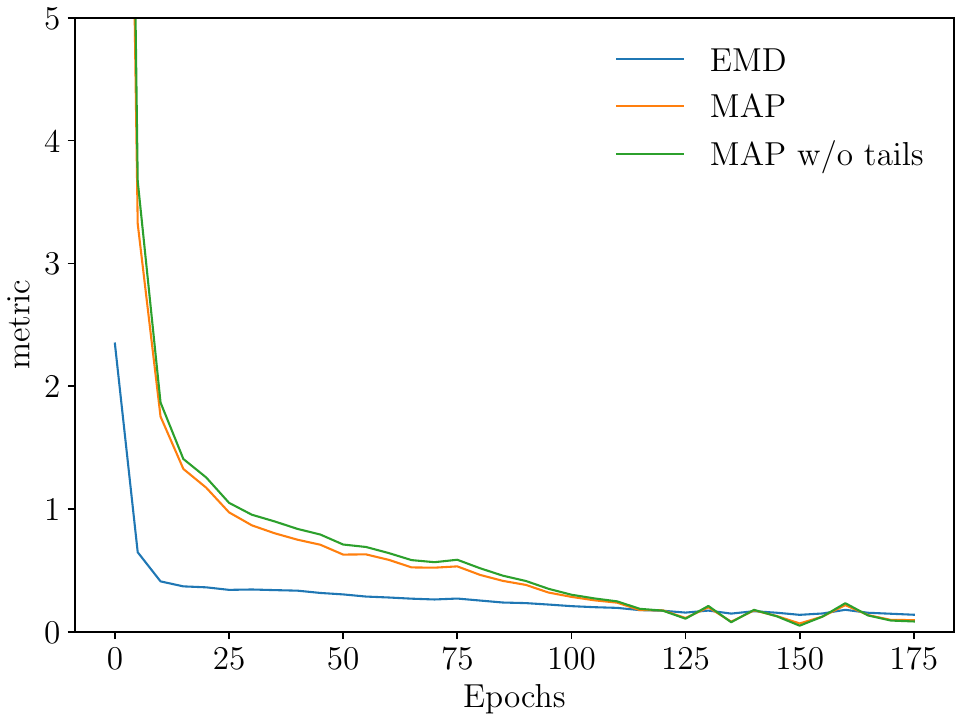}
    \caption{The EMD and MAP metrics evaluated during training of QE (left figure) and INC (right figure) models. We plot MAP$-\sqrt{2/\pi}$. Metrics are computed every $5$ epochs.}
    \label{fig:performance_EMD_MAP-QE}
\end{figure*}

Another way of checking the quality of the model is by estimating the so-called~\textit{pull}~\cite{DemortierEverythingYA,Alanazi:2020jod}. This technique compares the histograms of events generated by the \nuwro{} and the GAN.

The $i$-th component of the pull is given by 
\begin{equation}
    \text{pull}_i = \frac{n_{\nuwro{},i} - n_{gan,i}}{\sqrt{\sigma_{\nuwro{},i}^2 + \sigma_{gan,i}^2}},
    \label{eqn:pull}
\end{equation}
where $n_{\nuwro{},i}$ and $n_{gan,i}$ are the bin contents at the $i$-th bin, and $\sigma_{\nuwro{},i}^2=n_{\nuwro{},i}$ and $\sigma_{gan,i}^2=n_{gan,i}$ are statistical uncertainties. The bin contents of a histogram are expected to follow the Poisson distribution.
As the G event distribution gets closer to \nuwro{}'s, then Eq.~(\ref{eqn:pull}) defines a random variable that follows a standard normal distribution $N(0,1)$ for large $n_{\nuwro{},i}$ and $n_{gan,i}$. For a random variable $x\sim N(0,1)$, the expected value of $|x|$ is 
\begin{equation}
        \mathbb{E}_{x\sim N(0,1)}[\left|x\right|] = \sqrt{2/\pi} \approx 0.80.
    \end{equation}
To check the goodness of the model, we evaluate the mean of the absolute value of pulls
\begin{equation}
    \text{MAP}=\frac{1}{K}\sum_{i=1}^K\lvert\text{pull}_i\rvert,
    \label{eq:MAP}
\end{equation}
where $K$ is the number of bins that satisfy $n_{nuw,i}\neq0$ and $n_{gan,i}\ne0$. We checked that the MAP between two (normalized) \nuwro{} samples is  $\sim 0.8$.

The MAP metric is computed using histograms of the joint $E_{\mu}^\prime$ and $\theta^\prime$ distribution, as shown in Figs.~\ref{fig:pulls_map1}~and~\ref{fig:pulls_map2}. To reduce the bias coming from low statistics bins at distribution ``tails'', we compute the MAP for bins for which $n_{nuw,i}>5$ and $n_{gan,i}>5$.

The Wasserstein distance, or EMD~\cite{710701}, is another metric to monitor the quality of the models. It measures how different two given histograms are and how much ``work'' one must do to redistribute one histogram into another. We evaluate this metric to monitor the quality of the training and final models.  The EMD computed for two \nuwro{} samples is $\sim0.03$ - $0.04$.

\textit{A priori}, the MAP (\ref{eq:MAP}) and the EMD depend on the histogram binning. To minimize its impact, $\sim$one million events are generated with both \nuwro{} and the GAN, with a binning of $50\times50$ in $E_\nu^\prime$ and $\theta^\prime$.

In Fig.~\ref{fig:performance_EMD_MAP-QE}, we show how the EMD and MAP metrics change during the training of the QE and INC models. As can be noticed, after $100$ epochs, the optimization leads to a model for which the metrics begin to saturate.
We stop the optimization using the MAP and EMD metrics. The best model minimizes both.

\begin{figure*}[p]\centering
    \includegraphics[width=0.48\textwidth]{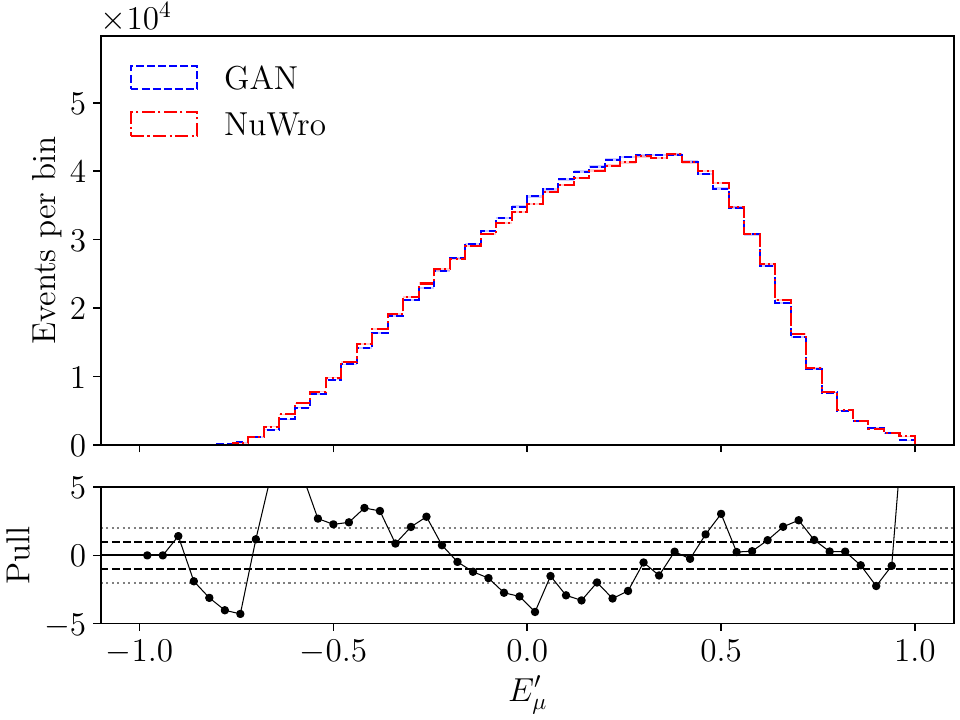} \ \ \
    \includegraphics[width=0.48\textwidth]{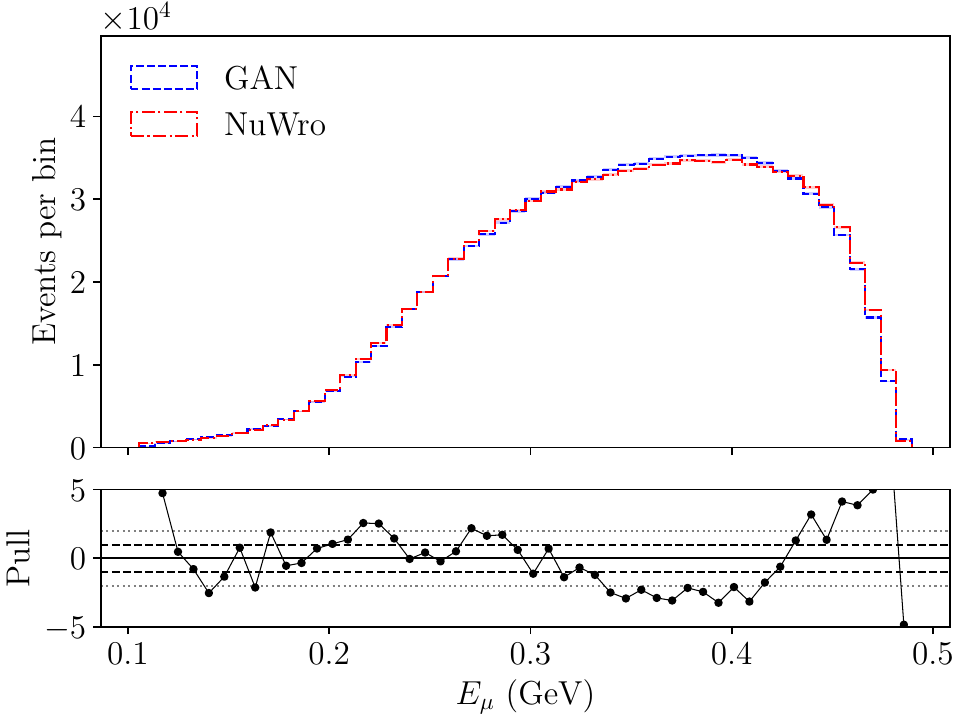} \\
    \includegraphics[width=0.48\textwidth]{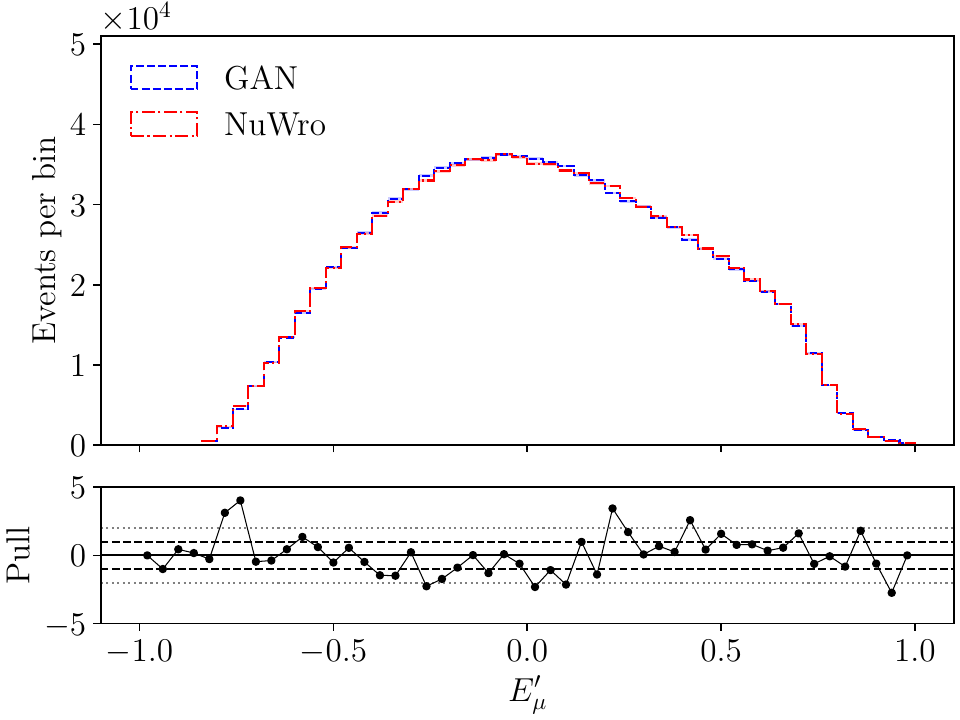} \ \ \
    \includegraphics[width=0.48\textwidth]{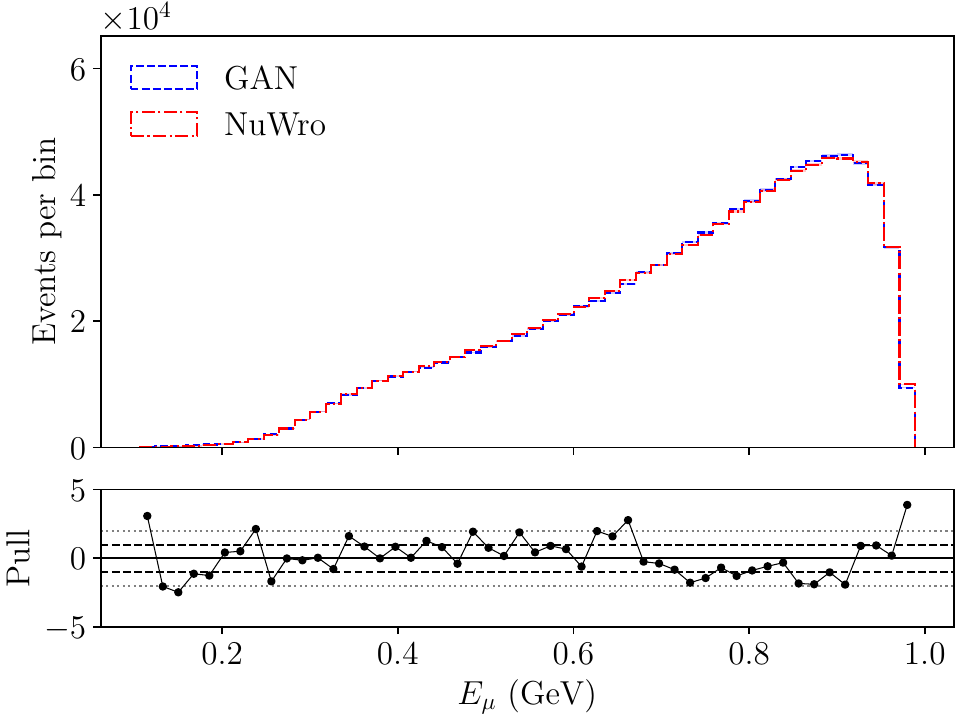} \\
    \includegraphics[width=0.48\textwidth]{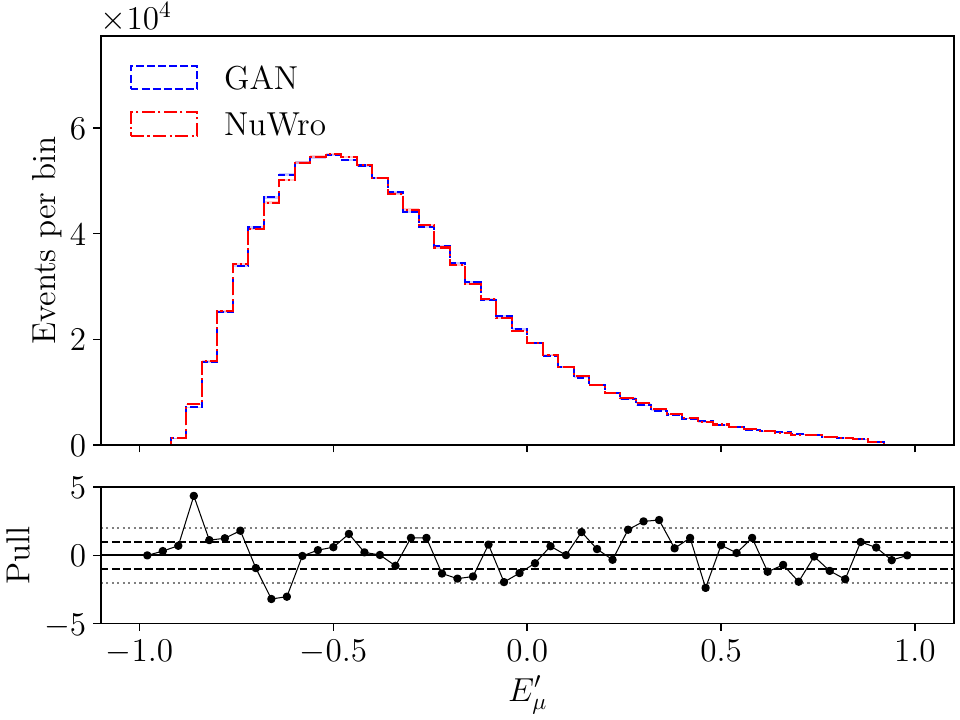} \ \ \
    \includegraphics[width=0.48\textwidth]{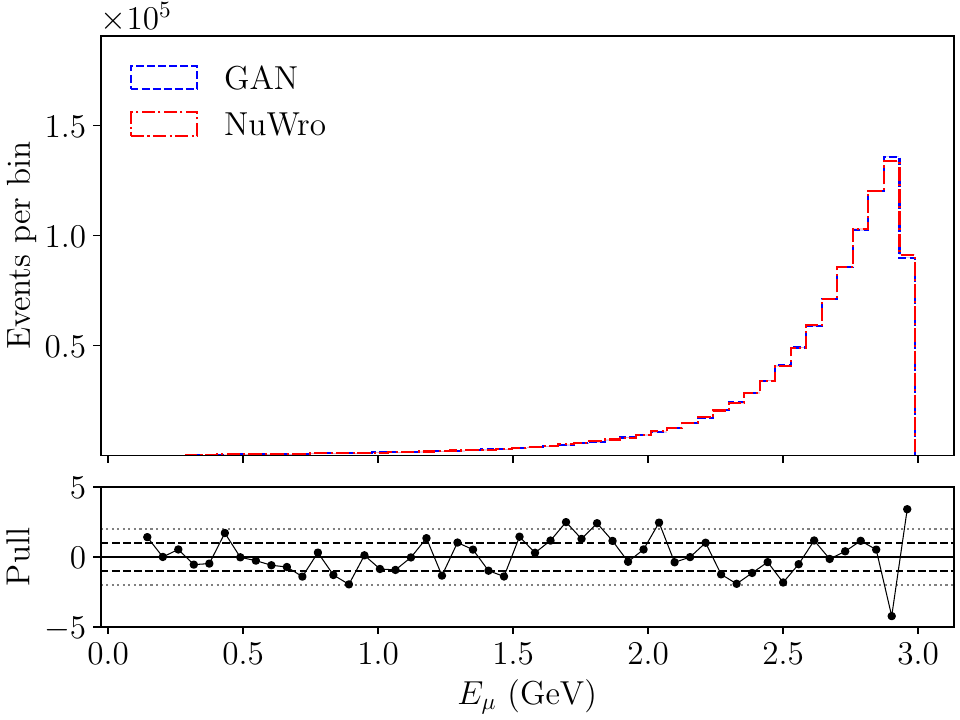}
    \caption{
    Event histograms, depending on $E_\mu^\prime$ (left column)  and $E_\mu$ (right column) generated by the G-QE and \nuwro{} test samples for neutrino energies $E_\nu =0.5$,~$1$,~and~$3$~GeV, from top to bottom rows, respectively.
    \label{fig:muonenergy-QE}}
\end{figure*}

\begin{figure*}[htbp]\centering
    \includegraphics[width=0.48\textwidth]{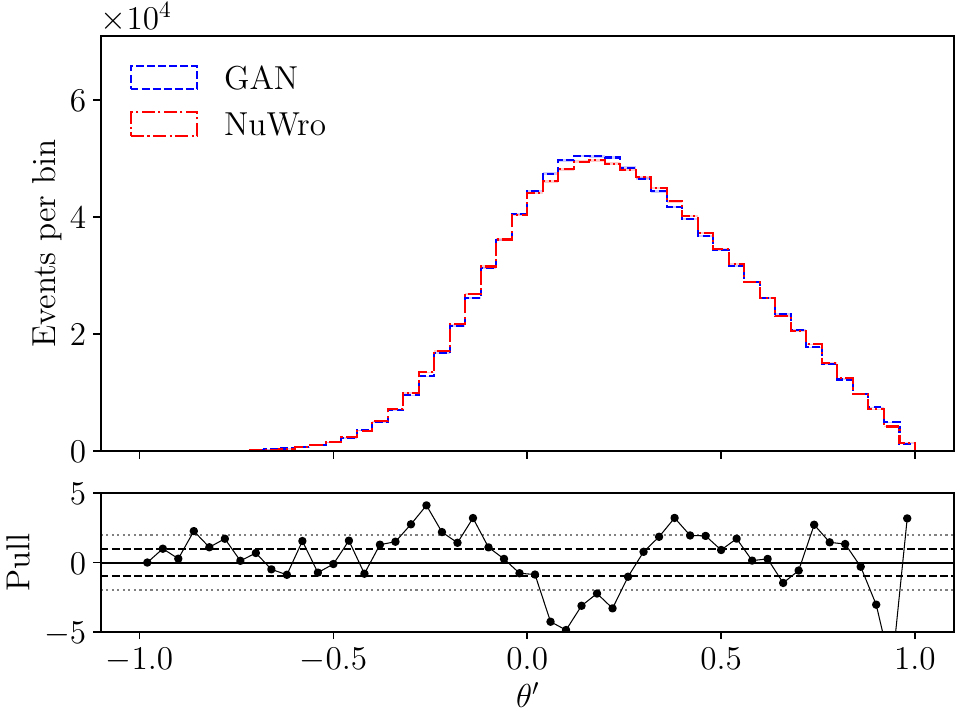} \ \ \
    \includegraphics[width=0.48\textwidth]{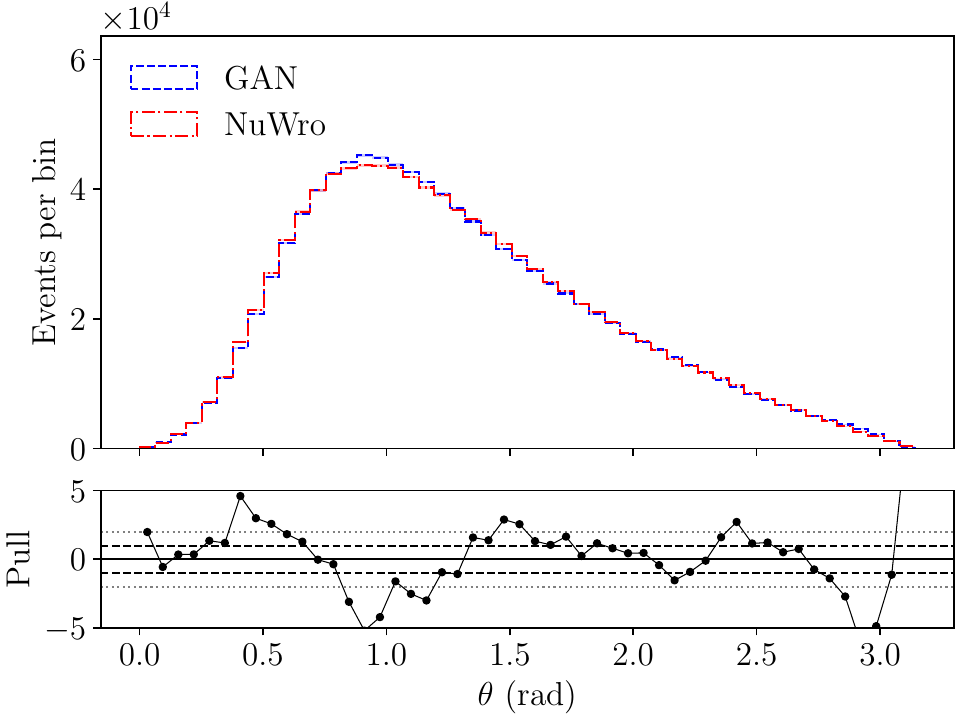} \\
    \includegraphics[width=0.48\textwidth]{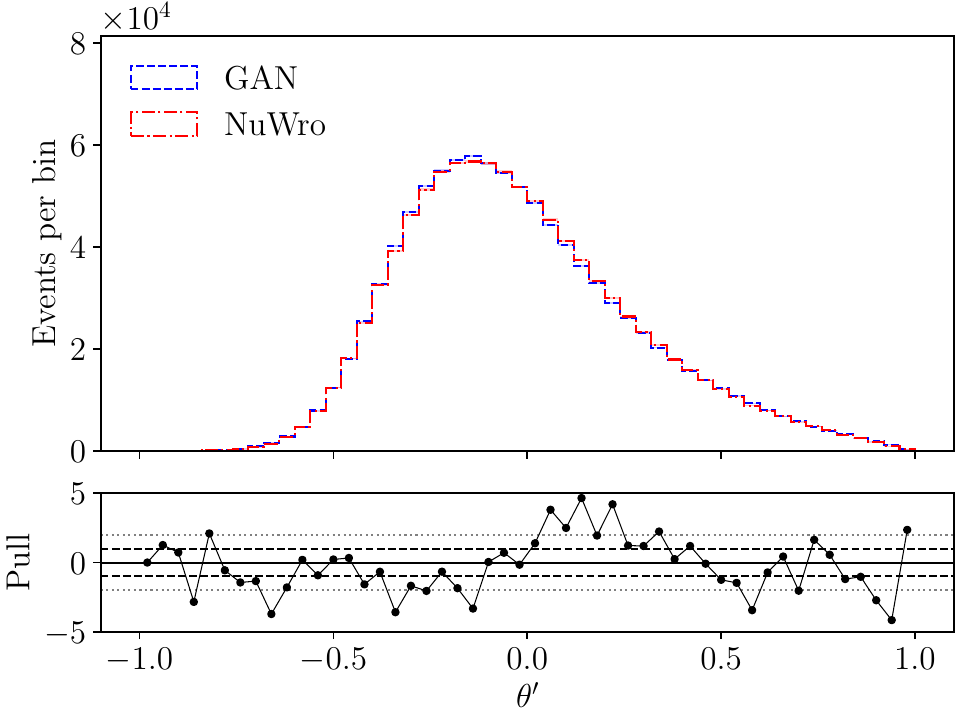} \ \ \
    \includegraphics[width=0.48\textwidth]{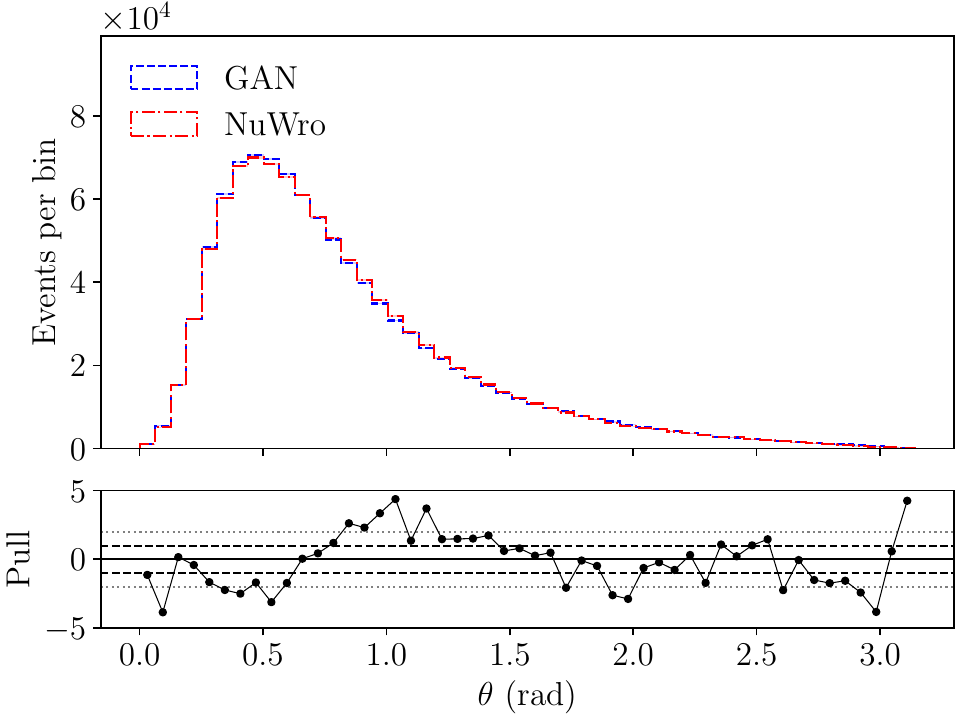} \\
    \includegraphics[width=0.48\textwidth]{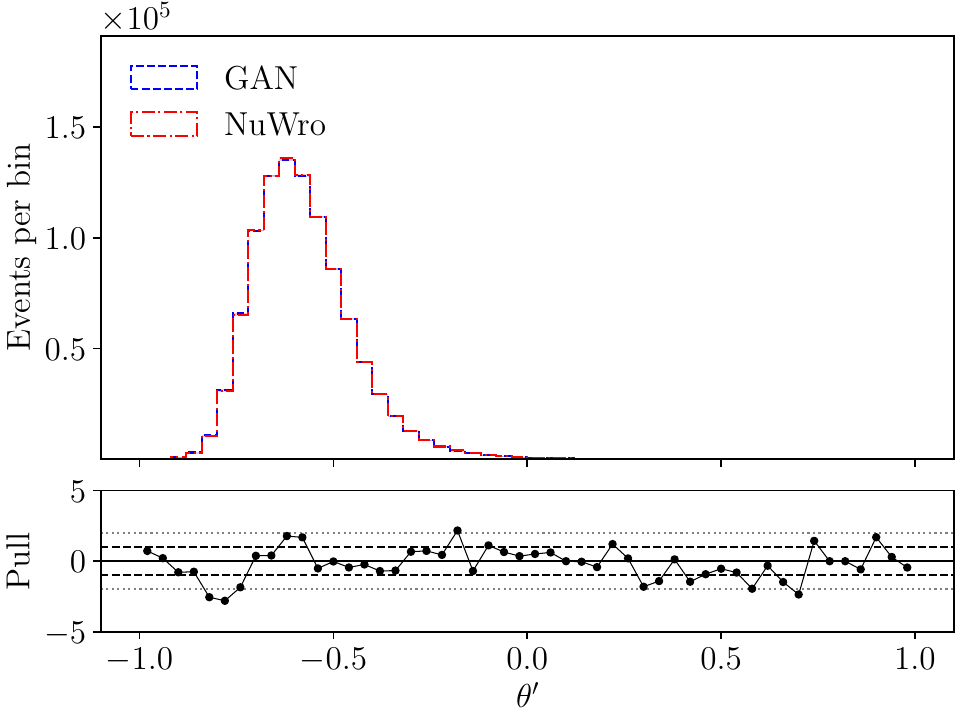}
    \includegraphics[width=0.48\textwidth]{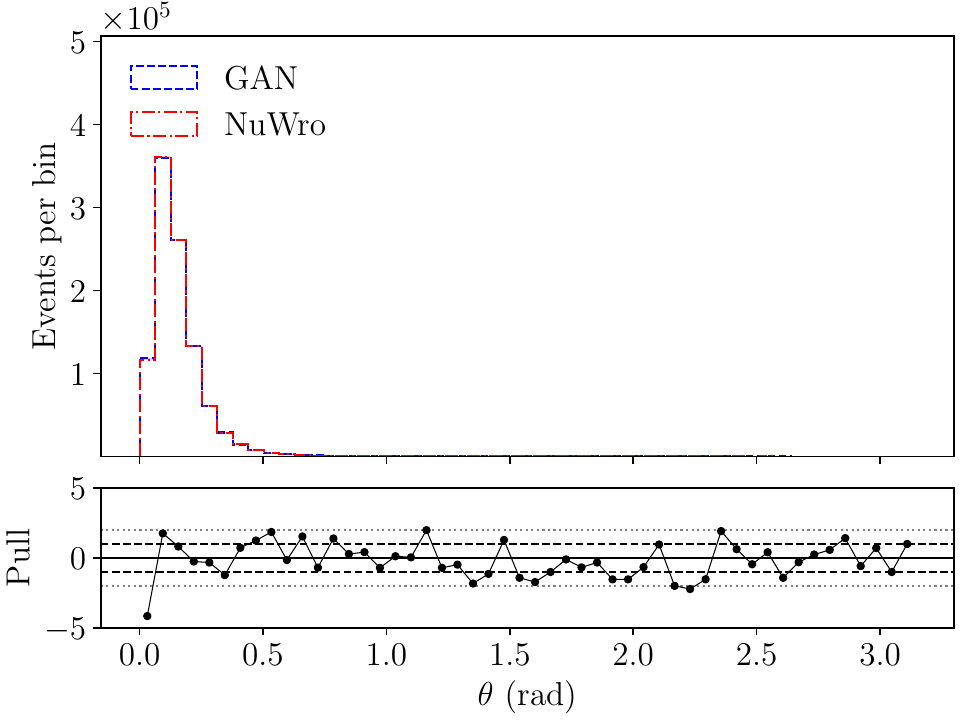}
    \caption{Event histograms, depending on $\theta^\prime$ (left column)  and $\theta$ (right column) generated by the G-QE and \nuwro{} test samples for neutrino energies~$=0.5$,~$1$,~ and~$5$~GeV, shown in top, middle, and bottom rows, respectively.
    \label{fig:theta-QE}}
\end{figure*}

\section{Results}
\label{Sec:Numerical}

In this section, we compare the predictions made by the GAN model with the test data. To assess the quality of our models, we generated several datasets using \nuwro{}. Each dataset was created for a specific value of neutrino energy, and we produced $1$~million events for each case. Additionally, we obtained test datasets (for QE and INC) similar to training, namely, one million events distributed uniformly in the energy range (300 MeV, 10 GeV).

In this section, we mainly discuss and compare GAN predictions with test datasets. In the appendix, we present the comparison between predictions of GAN and training datasets.

 We first present the numerical results for charged-current quasielastic scattering of muon neutrinos. 
\begin{table}[htbp]
    \caption{EMD and MAP (with and without tails) computed for the samples generated by the \nuwro{} test sample and the G-QE for fixed neutrino energy $E_\nu$ and randomly sampled from a uniform $E_\nu$ distribution (``All'' tag). \label{tab:QEmetrics}} 
\begin{ruledtabular}
    \begin{tabular}{c|rrr}
        $E_\nu$ & EMD  & MAP  & MAP w/o tails \\ \hline\hline
        500 MeV & 0.15 & 1.16 & 1.17          \\ \hline
        800 MeV & 0.14 & 1.06 & 1.06          \\ \hline
        1 GeV   & 0.14 & 1.02 & 1.00          \\ \hline
        2 GeV   & 0.14 & 1.01 & 0.98          \\ \hline
        3 GeV   & 0.13 & 0.98 & 0.97          \\ \hline
        5 GeV   & 0.13 & 0.92 & 0.90          \\ \hline
        7 GeV   & 0.15 & 1.05 & 1.02          \\ \hline
        9 GeV   & 0.11 & 1.03 & 0.99          \\ \hline
        All     & 0.11 & 0.90 & 0.84          
    \end{tabular}
    \end{ruledtabular}
\end{table}

\begin{figure*}[htpb]\centering
    \includegraphics[width=0.48\textwidth]{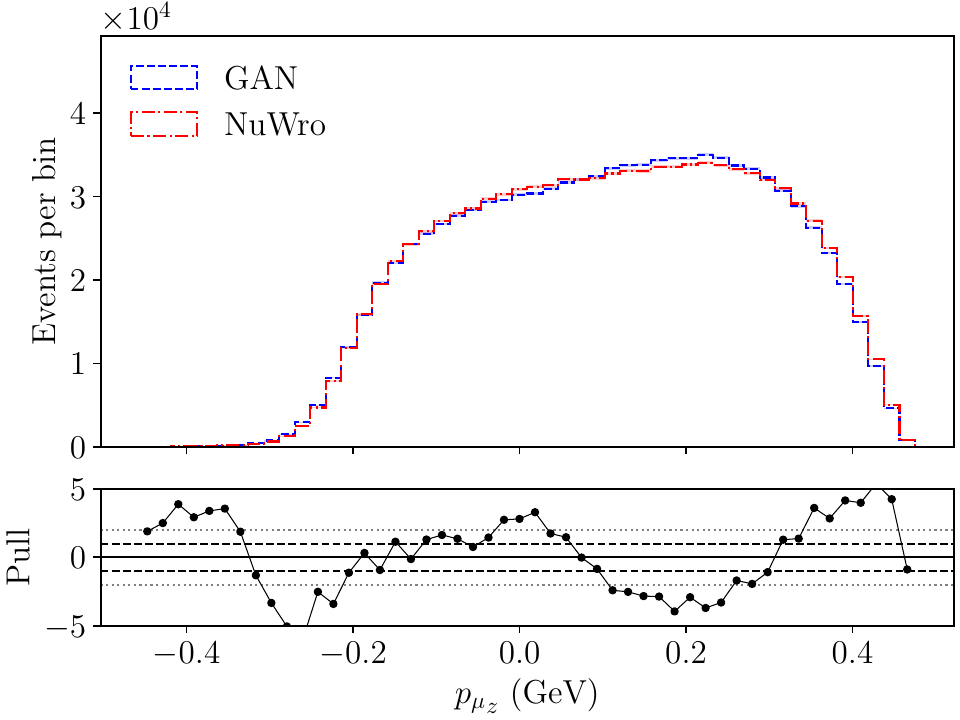} \ \ \
    \includegraphics[width=0.48\textwidth]{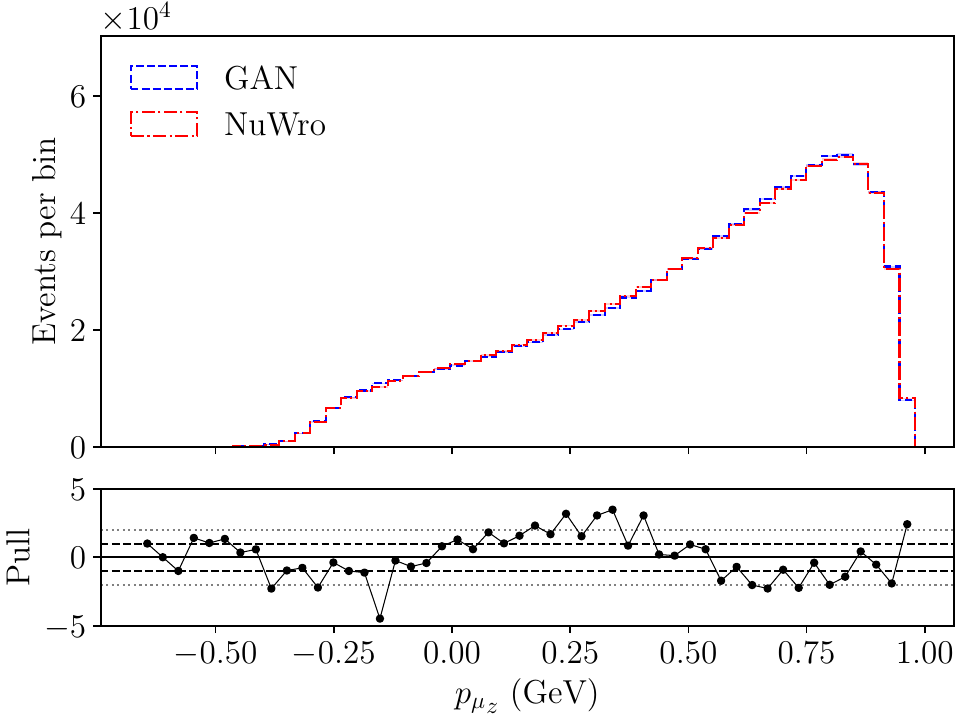} 
    \includegraphics[width=0.48\textwidth]{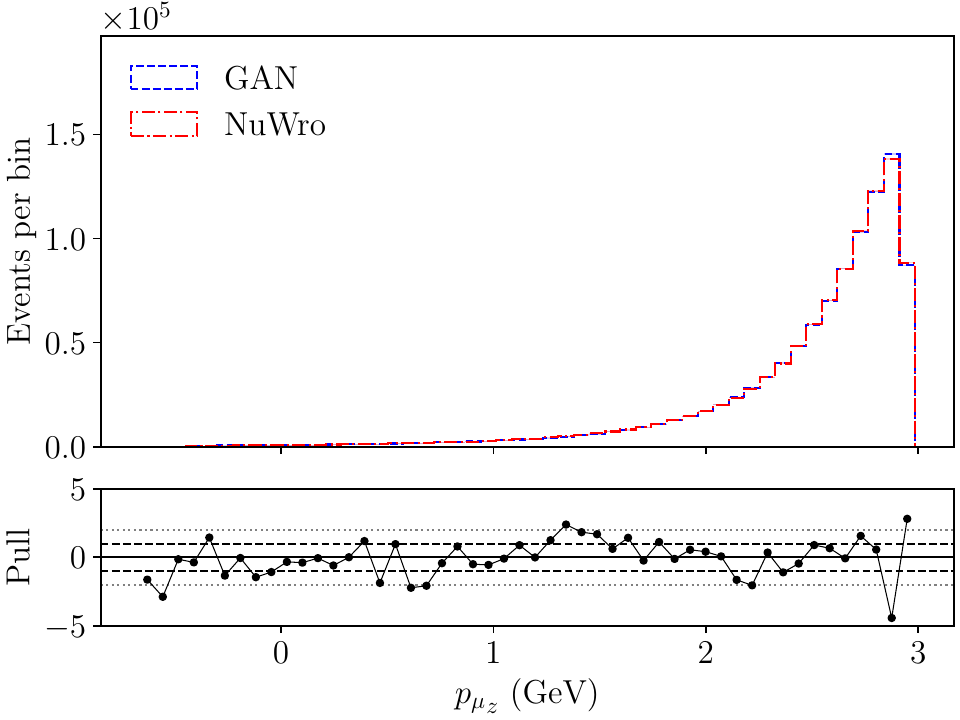}
    \caption{
    Event histograms, depending on $p_{\mu,z}$ generated by the G-QE and \nuwro{} test samples for neutrino energies~$0.5$,~$1$,~and~$3$~GeV,  shown in top left, top right, and bottom figures, respectively.
    \label{fig:pmuz-QE}}
\end{figure*}

The quality of the QE-GAN model and its dependence on neutrino energy is summarized in Table~\ref{tab:QEmetrics}. We provide the values of the EMD and MAP metrics computed for several values of neutrino energies, between $0.5$ and $9$~GeV, as well as for a dataset with energies distributed uniformly in the energy region that covers the same energy domain as the training dataset. The performance of the G-QE model is excellent in all the considered cases.

In Fig.~\ref{fig:pulls_map1}, we show the two-dimensional histograms and pull distribution of events generated by the \nuwro{} and G-QE that are uniformly distributed in energies from $300$~MeV to $10$~GeV. 
In both GAN- and \nuwro{}-generated samples, we clearly see the region where the events are concentrated. It corresponds to QE peak. The pulls per bin resemble statistical fluctuations, and their distribution approaches the Gaussian shape.

The G-QE model generates event histograms that closely match those produced by the \nuwro{}. This is demonstrated in Fig.~\ref{fig:muonenergy-QE}, where we present event histograms for neutrino energies of $0.5$,~$1$,~and $3$~GeV. The histograms are shown in terms of $E'_\mu$  and muon energy, along with an estimate of the \textit{pulls}. Notably, the distribution of the QE events in muon energy exhibits a single peak.

Fig.~\ref{fig:theta-QE} presents the dependence on $\theta'$ and $\theta$ for neutrino energies of $0.5$,~$1$,~and~$5$~GeV. Furthermore, we derive the longitudinal momentum as a composite variable (\ref{Eq:pmuz}) and show its event distribution in Fig.~\ref{fig:pmuz-QE}. The agreement between the GAN and the \nuwro{} generated results is excellent. Note that for the comparison, we generated a new set of the \nuwro{} events that were not used during training for all these tests.

\begin{figure*}[p]\centering
    \includegraphics[width=0.48\textwidth]{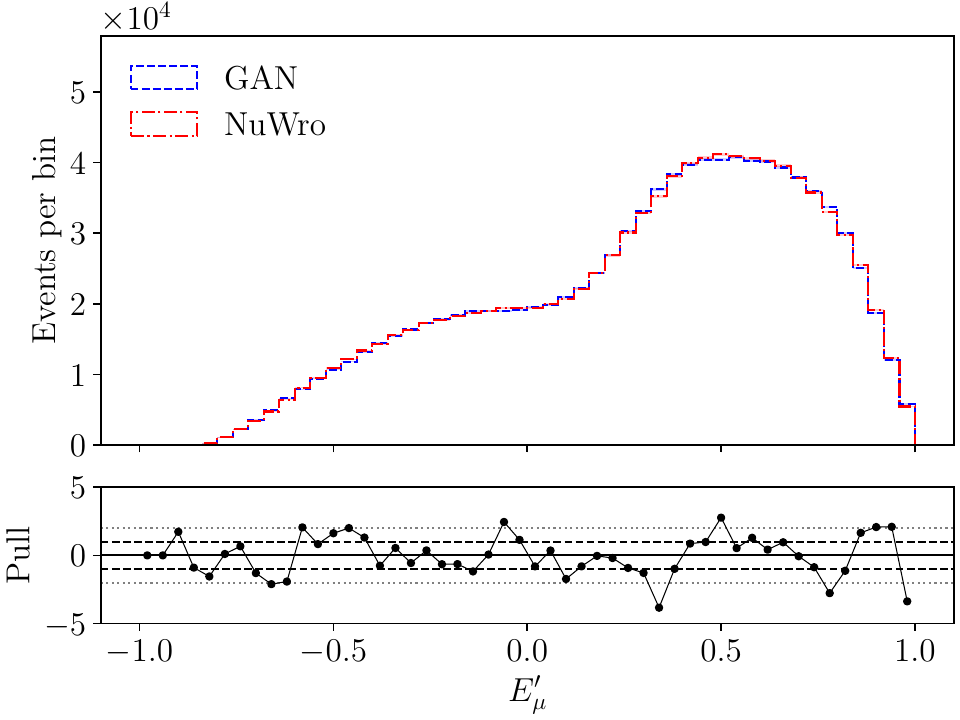} \ \ \
    \includegraphics[width=0.48\textwidth]{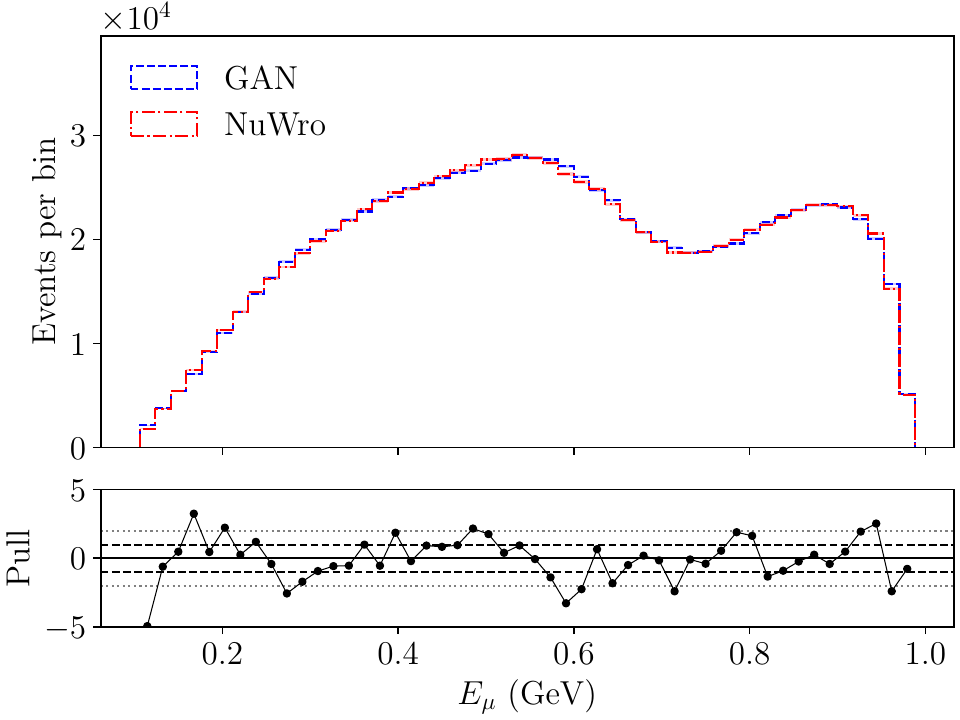} \\
    \includegraphics[width=0.48\textwidth]{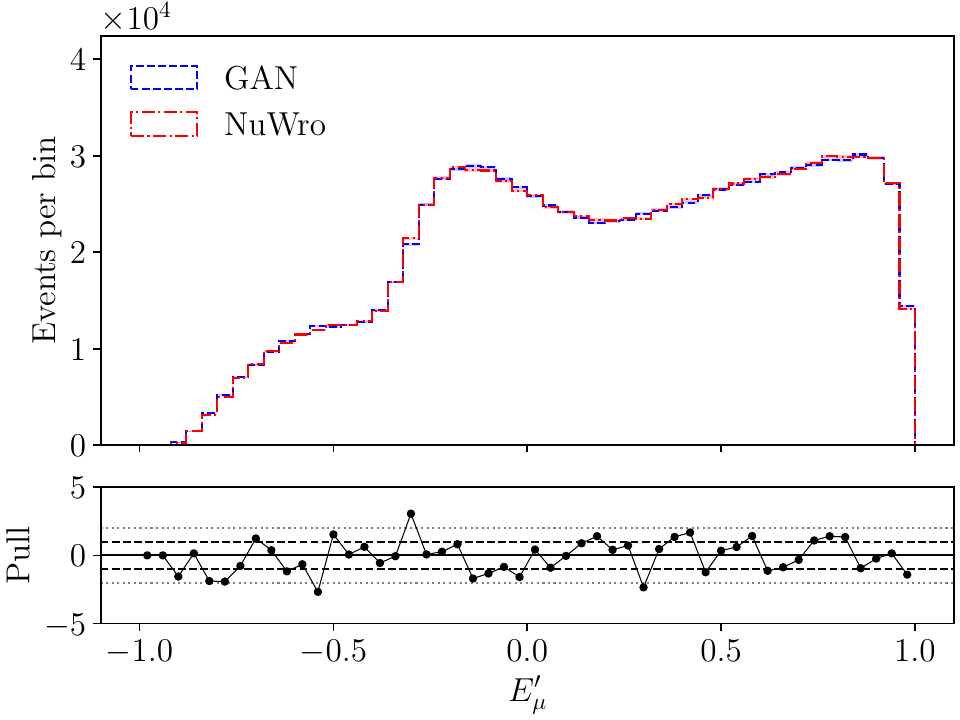} \ \ \
    \includegraphics[width=0.48\textwidth]{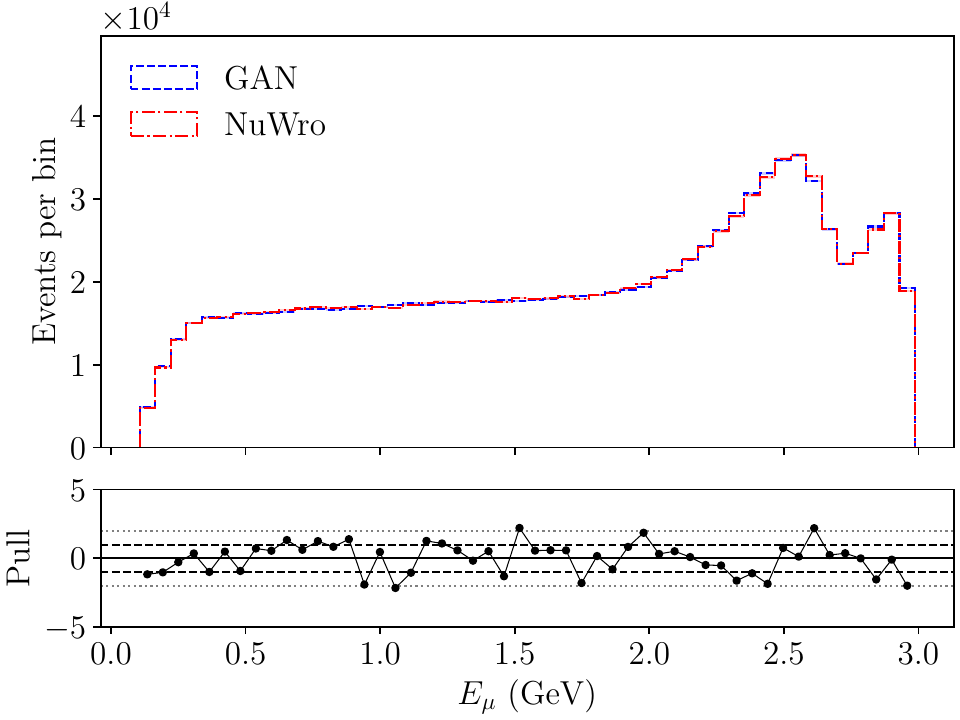} \\
    \includegraphics[width=0.48\textwidth]{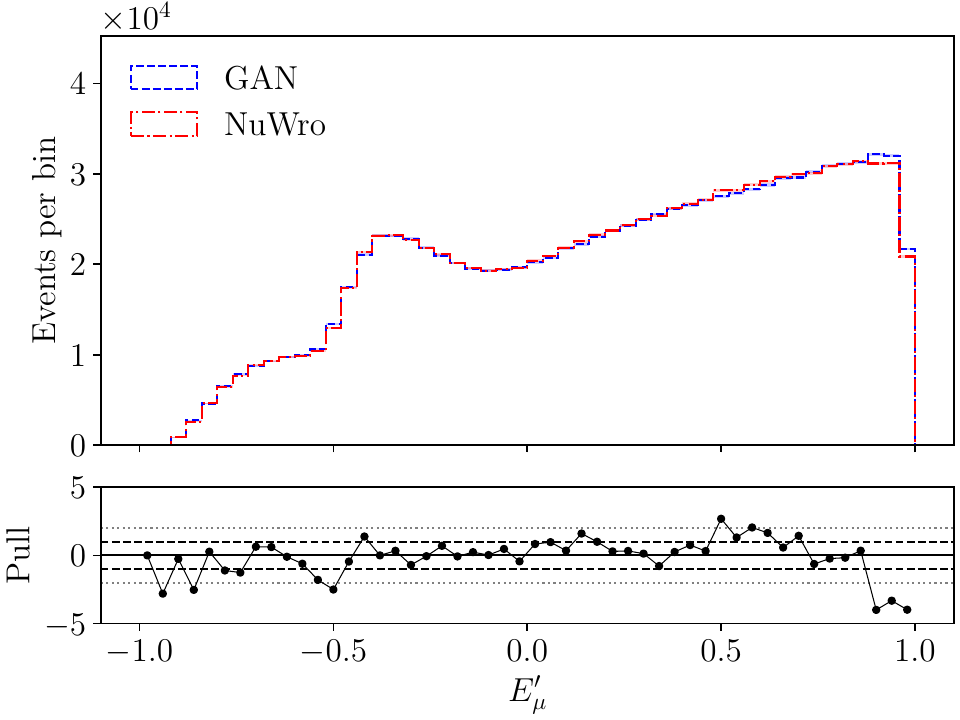} \ \ \
    \includegraphics[width=0.48\textwidth]{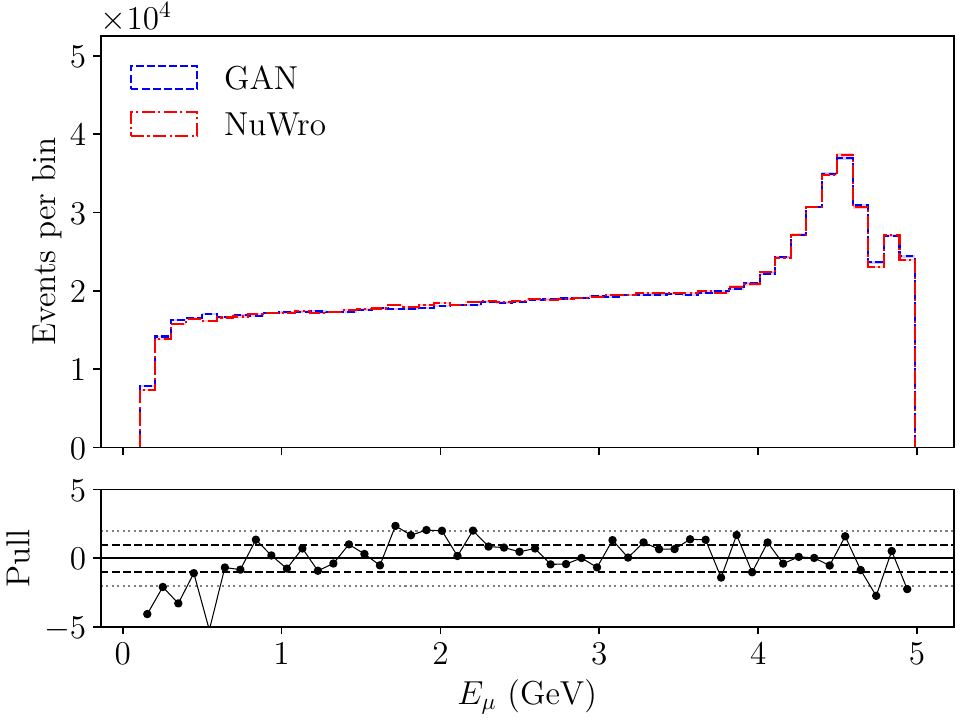}
    \caption{
    Event histograms, depending on $E_\mu^\prime$ (left column)  and $E_\mu$ (right column) generated by the G-INC and the \nuwro{} test samples for neutrino energies~$=1$, $3$, and $5$~GeV, shown in top, middle and bottom rows, respectively.
    \label{fig:incmuonenergy}}
\end{figure*}

Let us now discuss the INC model.  
In Fig.~\ref{fig:pulls_map2}, we show the two-dimensional histograms and pull distribution of INC events generated by the \nuwro{} and G-INC that are uniformly distributed in energies from $300$~MeV to $10$~GeV. Again, a good agreement is achieved between the \nuwro{} and G-INC generator. Both GAN and \nuwro{} reproduce the QE and $\Delta$ resonance peaks, see two accumulations of events in Fig.~\ref{fig:pulls_map2} (top panels).
In Fig.~\ref{fig:incmuonenergy}, we plot the histograms of events generated by the G-INC and the \nuwro{} MC generator for neutrino energies $1$,~$3$~and~$5$~GeV for both $E'_\mu$ and $E_\mu$. 

\begin{figure*}[htpb]\centering
    \includegraphics[width=0.48\textwidth]{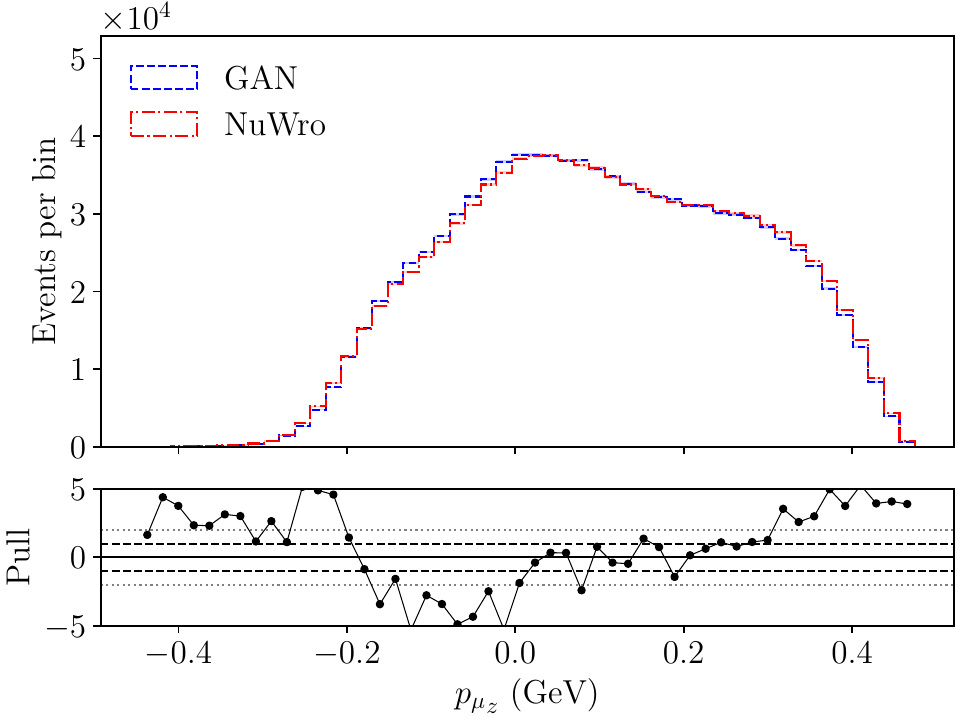} \ \ \
    \includegraphics[width=0.48\textwidth]{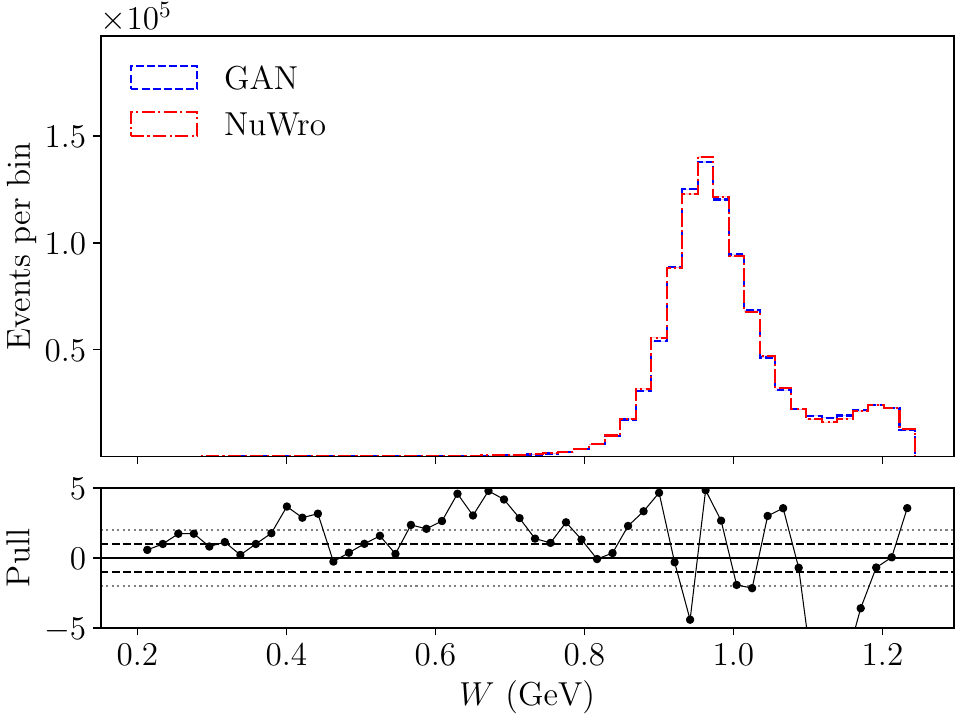} \\
    \includegraphics[width=0.48\textwidth]{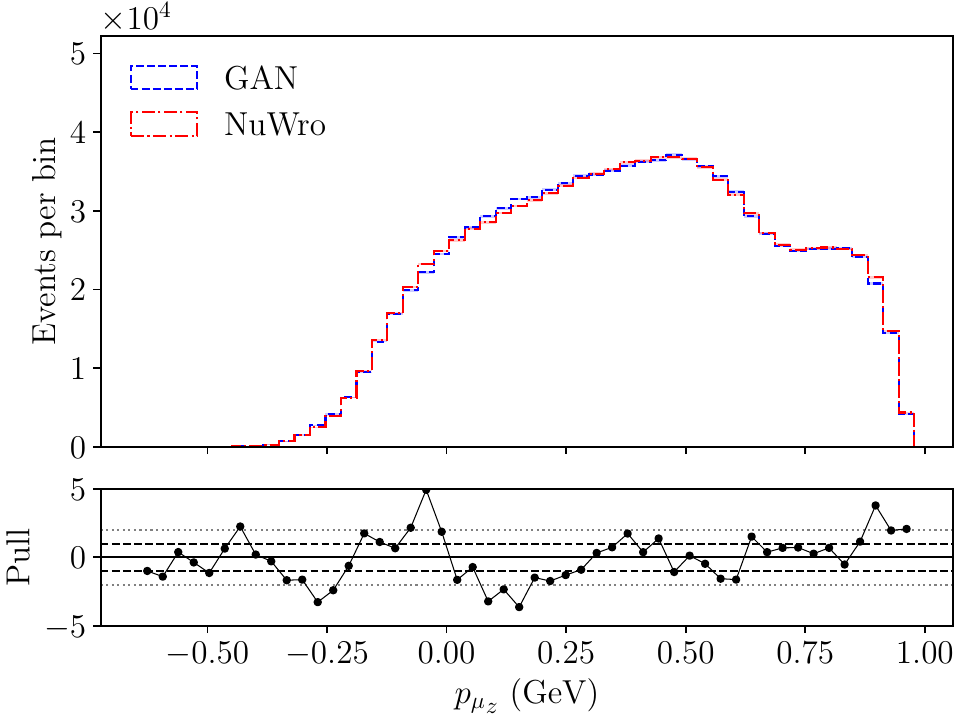} \ \ \ 
    \includegraphics[width=0.48\textwidth]{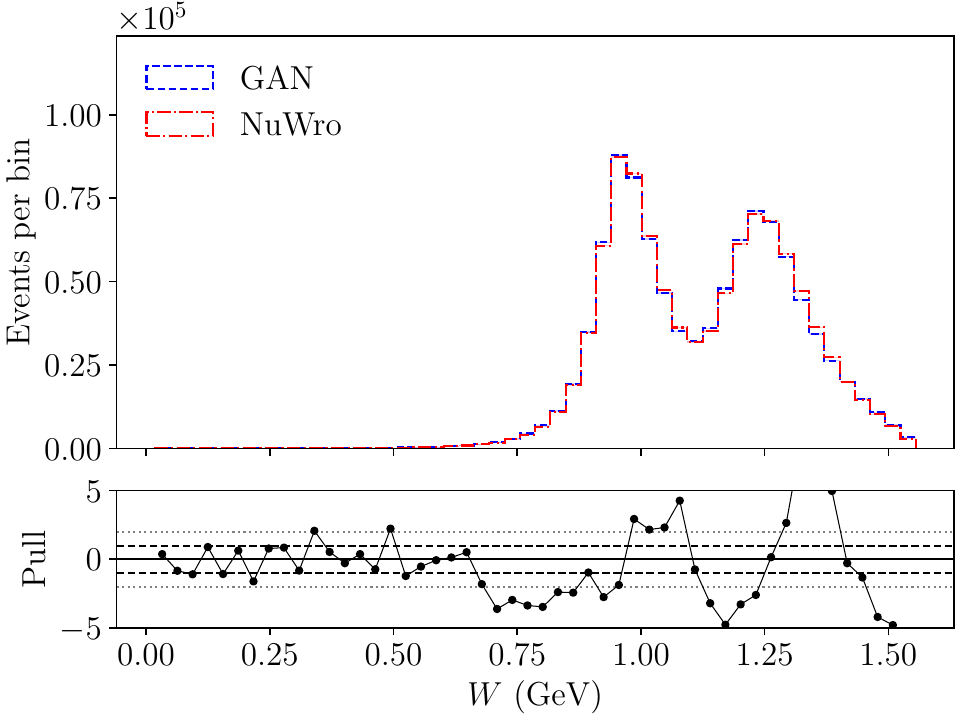} \\
    \includegraphics[width=0.48\textwidth]{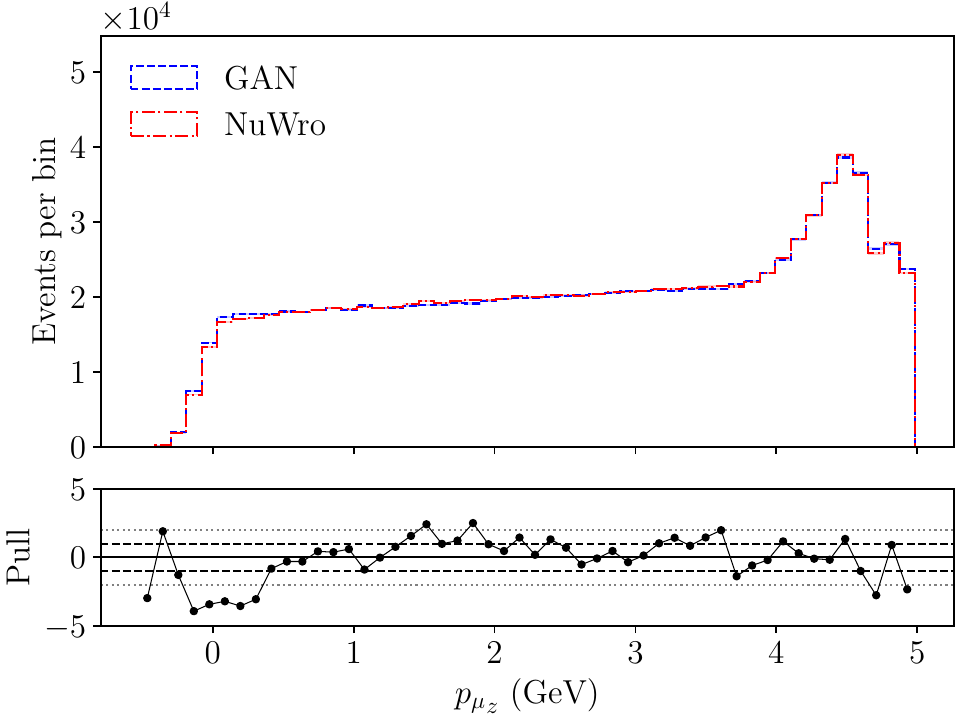} \ \ \
    \includegraphics[width=0.48\textwidth]{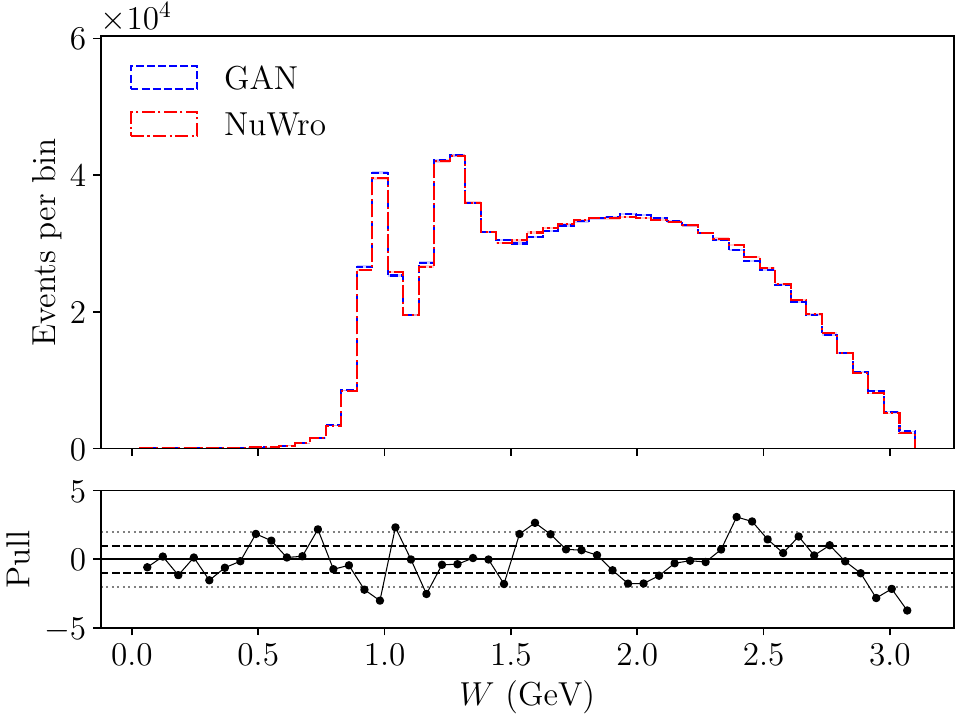}
    \caption{Event histograms, depending on $p_{\mu,z}$ (left column)  and hadronic invariant mass $W$ (right column) generated by the G-INC and the \nuwro{} test samples for neutrino energies $E_\nu =0.5$, $1$, and $5$~GeV, shown in top, middle and bottom rows, respectively. \label{fig:incmuonpz}}
\end{figure*}

To test if our GAN model can reproduce the QE and $\Delta(1232)$ resonance peaks equally well, let us now consider event distributions in hadronic invariant mass $W$ defined as
\begin{equation}
    W = \sqrt{M_N^2+2 \omega M_N - Q^2},
    \label{Eq:massW}
\end{equation}
where $\omega = E_\nu - E_\mu$ is the  energy transferred to the nucleus, $M_N$ the average nucleon mass of the nucleon, and $Q^2$ the four-momentum transfer
\begin{equation}
    Q^2 = 2 E_\nu (E_{\mu} - p_{\mu,z}) - m_\mu^2.
\end{equation}

In Fig.~\ref{fig:incmuonpz}, we present the distributions of events as a function of longitudinal muon momentum \( p_{\mu,z} \) and hadronic invariant mass \( W \). We observe good agreement between the results from \nuwro{} and G-INC for these variables, both in the quasielastic (QE) region and at the peak of the \( \Delta(1232) \) resonance. This consistency extends to the onset of deep inelastic (DIS) region in the histograms for \( E_\nu = 5 \) GeV.
\begin{figure}[htbp]
	\includegraphics[width=0.45\textwidth]{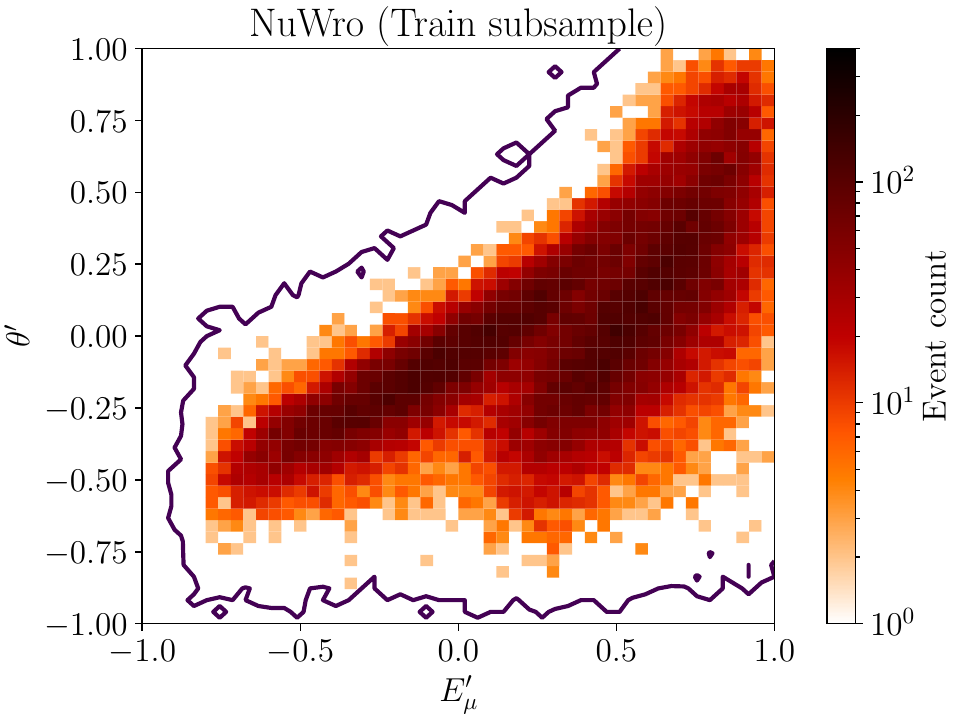}
	\includegraphics[width=0.45\textwidth]{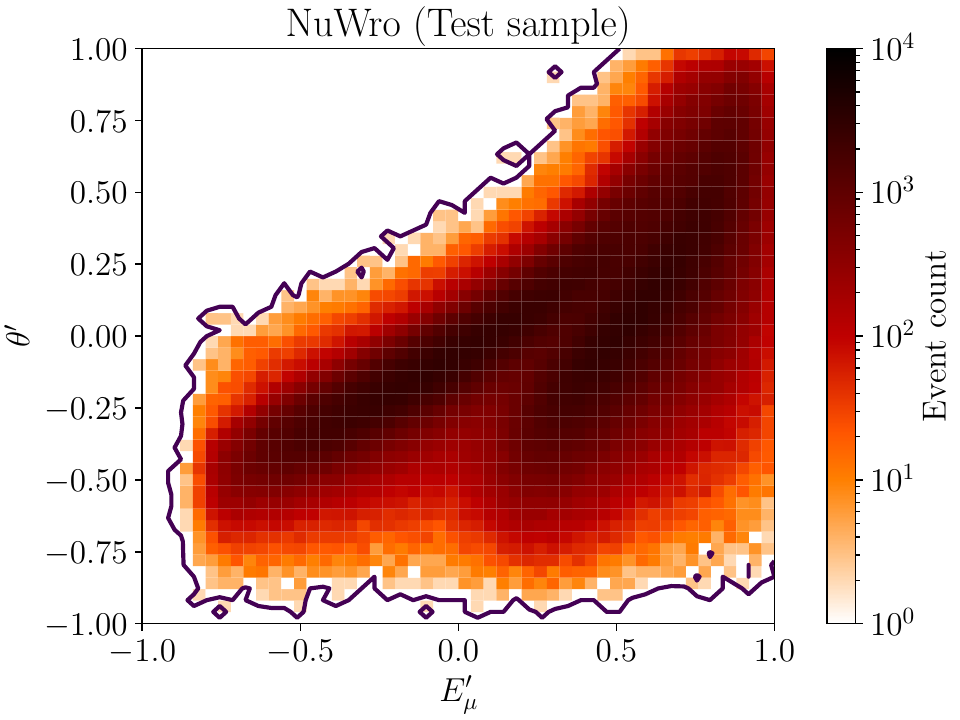}
	\caption{Two-dimensional inclusive distributions of \nuwro{} events in $E_\mu^\prime$ and $\theta^\prime$ variables. Top: the training dataset for $E_\nu$ ranging from 950 MeV to 1050 MeV ($\sim40$k events). Bottom: Testing dataset of one million events for $1$ GeV neutrino energy. Both figures show the approximate outline (based on histogram binning) of the $1$~GeV dataset distribution. The training dataset does not cover the whole kinematic range of the \nuwro{} generator for neutrino energy around 1~GeV.
		\label{Fig:2dTrainTest}}
\end{figure}

\begin{table}[htbp]
    \caption{Caption the same as in Table \protect\ref{tab:QEmetrics} but for INC model.} \label{tab:Inlusivemetrics} 
\begin{ruledtabular}
    \begin{tabular}{c|rrr}
        $E_\nu$ & EMD  & MAP  & MAP w/o tails \\ \hline\hline
        500 MeV & 0.16 & 1.08 & 1.07          \\ \hline
        800 MeV & 0.16 & 1.09 & 1.09          \\ \hline
        1 GeV   & 0.15 & 1.07 & 1.07          \\ \hline
        2 GeV   & 0.15 & 1.02 & 1.00          \\ \hline
        3 GeV   & 0.14 & 0.94 & 0.92          \\ \hline
        5 GeV   & 0.14 & 0.96 & 0.93          \\ \hline
        7 GeV   & 0.16 & 1.01 & 0.99          \\ \hline
        9 GeV   & 0.15 & 1.03 & 1.02          \\ \hline
        All     & 0.13 & 0.89 & 0.86
    \end{tabular}
    \end{ruledtabular}
\end{table}

The numerical results for the INC model are summarized in Table~\ref{tab:Inlusivemetrics}, where we present metrics EMD and MAP computed for test data sets generated for several values of the neutrino energies. The quality of performance of the G-INC model is as good as for the G-QE.

It's important to note that the test dataset covers a wider range of lepton kinematics than the training data. In Fig.~\ref{Fig:2dTrainTest}, we display a histogram of the test data (INC) for neutrino energy at 1 GeV, alongside the training data for neutrino energies between $950$~MeV and $1050$~MeV. This comparison reveals that there are areas of lepton kinematics that are not represented in the training dataset but still are effectively reproduced by the GAN.

\section{Summary}
\label{Sec:Summary}

This article discusses the development of generative adversarial network models for simulating quasielastic and inclusive neutrino-nucleus scattering. The first type of interaction plays a pivotal role in the oscillation analyses carried out by the T2K and Hyper-Kamiokande experiments, and the other is important for the DUNE experiment.

Our model approach focuses on a simplified scenario in which the models, for neutrino energies ranging from $300$~MeV to $10$~GeV, predict muon kinematic variables—namely, its energy and scattering angle. We consider various kinematic distributions of the charged lepton (for neutrino energy from $300$~MeV to $10$~GeV). The models we present successfully reproduce the peak structure of the data distributions. 

Once these models are developed, they generate events faster than ``standard'' generators\footnote{Generating 1 million events on a single-core CPU takes approximately 12 minutes with NuWro (full simulation), and around 28 seconds with GAN. When GAN predictions are made on a typical gaming GPU (RTX 4080), the time is reduced to approximately 3 seconds.}. We also anticipate that these models can be adapted to more realistic scenarios after retraining them on experimental data. Essentially, they can serve as pre-trained models that can be fine-tuned for specific applications.

Our study opens the door for future developments, including considering complete event topologies and realistic neutrino fluxes. Furthermore, these deep neural network models can be repurposed to simulate related processes by utilizing advanced deep learning techniques such as transfer learning~\cite{graczyk2024electronnucleuscrosssectionstransfer}. These studies are underway. Combining these ideas will hopefully lead to MC simulation tools that can improve their predictive precision by learning from new experimental measurements and theoretical considerations.

\begin{acknowledgments}
This research is partly (K.M.G., A.M.A., J.T.S.) or fully (B.E.K., J.L.B, H.P., R.D.B.) supported by the Na{\-}tional Science Centre under grant UMO-2021/41/B/ST2/ 02778.
J.L.B. and J.T.S. are also supported by Polish Ministry of Science grant 2022/WK/15.
K.M.G is partly supported by the ``Excellence Initiative – Research University" for the years 2020-2026 at the University of Wroc\l aw.
\end{acknowledgments}

\appendix

\section{GAN versus training data}

\begin{figure*}[htpb]\centering
    \includegraphics[width=0.48\textwidth]{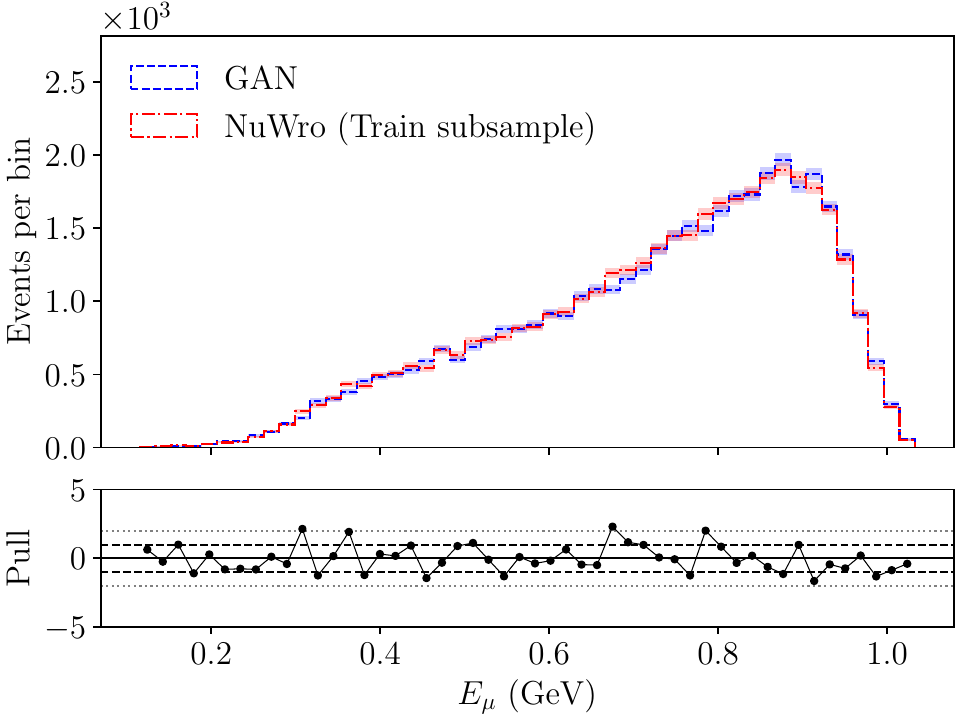} \ \ \
    \includegraphics[width=0.48\textwidth]{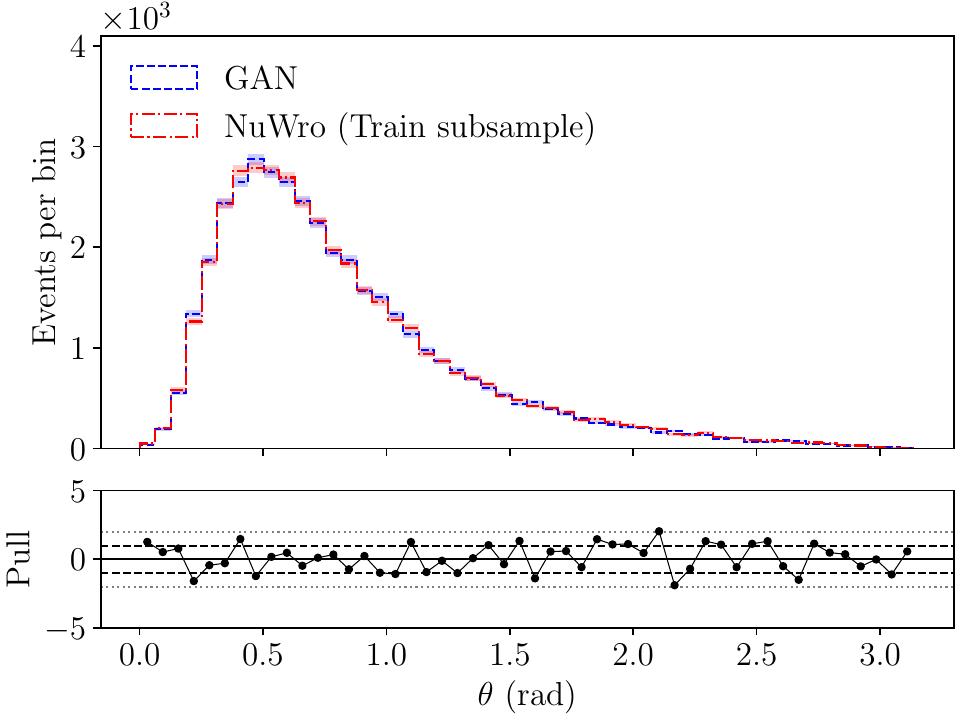} \\
    \includegraphics[width=0.48\textwidth]{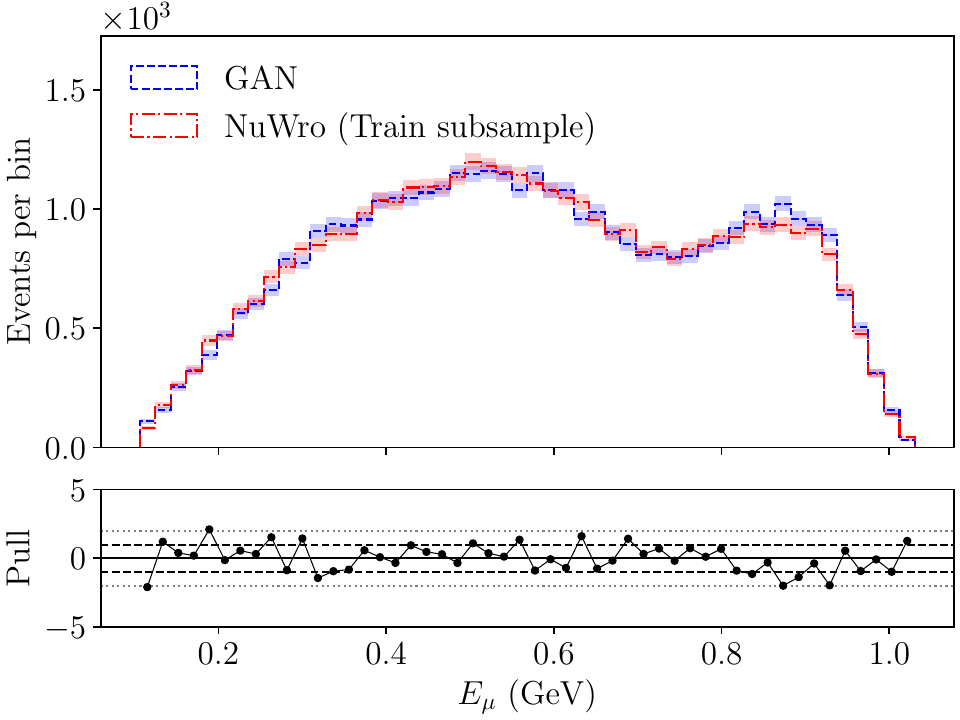} \ \ \ 
    \includegraphics[width=0.48\textwidth]{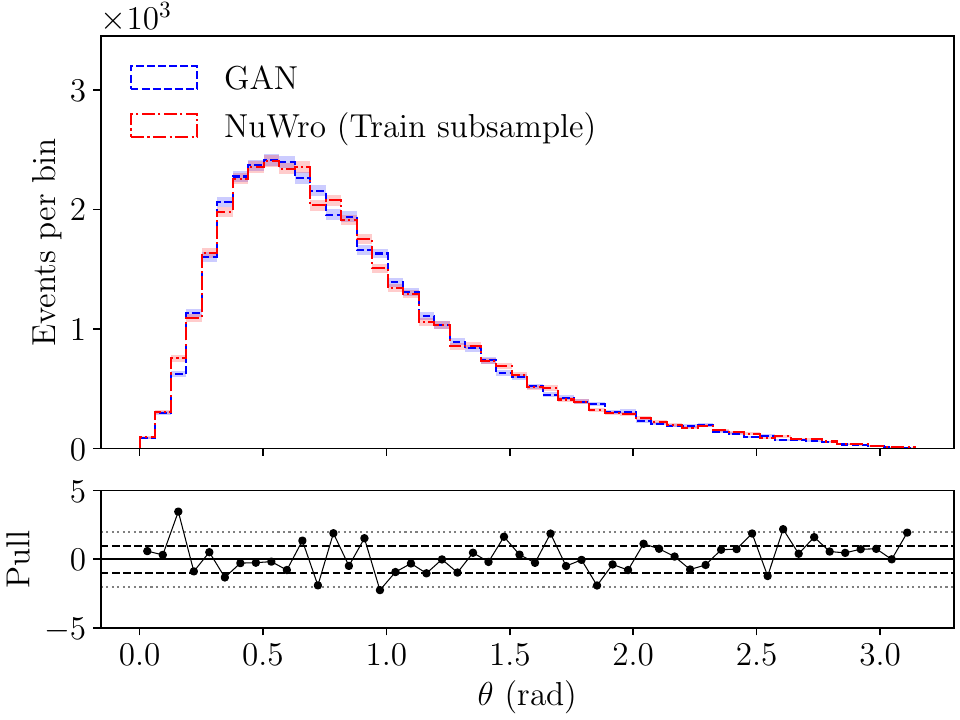}
    \caption{Event histograms of the training subsamples and their GAN counterpart. Top row: QE histograms. Bottom row: INC histograms. Left column: $E_\mu$ distributions. Right column: $\theta$ distributions. \label{fig:trainsubsamples}}
\end{figure*}

In Fig.~\ref{fig:trainsubsamples} we present comparison of the GAN predictions against training dataset. The training and GAN generated data are for neutrino energies larger than $950$~MeV and lower than  $1050$ MeV.

\bibliographystyle{apsrev4-2}
\bibliography{dnn,bibmoje,physics,bibdata}

\end{document}